\documentclass[traditabstract]{aa}

\usepackage{natbib,twoopt}
\usepackage[breaklinks=true]{hyperref} 
\bibpunct{(}{)}{;}{a}{}{,}             
\makeatletter
  \newcommandtwoopt{\citeads}[3][][]{\href{http://adsabs.harvard.edu/abs/#3}%
    {\def\hyper@linkstart##1##2{}%
     \let\hyper@linkend\@empty\citealp[#1][#2]{#3}}}
  \newcommandtwoopt{\citepads}[3][][]{\href{http://adsabs.harvard.edu/abs/#3}%
    {\def\hyper@linkstart##1##2{}%
     \let\hyper@linkend\@empty\citep[#1][#2]{#3}}}
  \newcommandtwoopt{\citetads}[3][][]{\href{http://adsabs.harvard.edu/abs/#3}%
    {\def\hyper@linkstart##1##2{}%
     \let\hyper@linkend\@empty\citet[#1][#2]{#3}}}
  \newcommandtwoopt{\citeyearads}[3][][]%
    {\href{http://adsabs.harvard.edu/abs/#3}
    {\def\hyper@linkstart##1##2{}%
     \let\hyper@linkend\@empty\citeyear[#1][#2]{#3}}}
\makeatother

\usepackage{graphicx}
\usepackage{txfonts}
\usepackage{amsmath}
\usepackage{siunitx}

\usepackage{hyperref}
\hypersetup{colorlinks=true,
            citecolor=blue,
            linkcolor=blue}
\usepackage{color}

\def \vv#1{\mathbf{#1}}

\def \lambdai {\lambda_1}
\def \lambdaj {\lambda_2}
\def \Lambdai {\Lambda_1}
\def \Lambdaj {\Lambda_2}

\def \psia {\varphi}
\def \Psia {\Phi}
\def \psii {\psia_1}
\def \psij {\psia_2}
\def \Psii {\Psia_1}
\def \Psij {\Psia_2}

\def \angrotU {\omega}

\def \xi    {x_1}

\def \yi    {y_1}
\def \ybi   {\overline{y}_1}
\def \yj    {y_2}
\def \ybj   {\overline{y}_2}
\def \yk    {y_k}
\def \ybk   {\overline{y}_k}

\def \Ka    {\mathcal{K}_a}
\def \Kb    {\mathcal{K}_b}
\def \Oc    {\mathcal{O}_a}
\def \Od    {\mathcal{O}_b}

\def \Sa    {\mathcal{S}_a}
\def \Se    {\mathcal{S}_b}
\def \Sf    {\mathcal{S}_c}
\def \Sg    {\mathcal{S}_7}
\def \Sg    {\mathcal{S}_d}
\def \Rd    {\mathcal{R}_a}
\def \Re    {\mathcal{R}_b}
\def \Rf    {\mathcal{R}_c}

\def \Ak    {\mathcal{A}_k}
\def \Bk    {\mathcal{B}_k}
\def \Ck    {\mathcal{C}_k}
\def \Jk    {\mathcal{J}_k}

\def \Ham {\mathcal{H}}
\def \avHam {\bar \Ham}
\def \HamS {\Ham_0}
\def \dtU {\tau}
\def \inc {I}
\def \ep {{\rm e}}
\def \ii {{\rm i}}
\def \degree {^{\circ}}
\def \At {A}

\def \mmi {\overline{n}_1}     
\def \mmj {\overline{n}_2}     

\def \vs {\vv{s}}
\def \vr {\vv{r}}
\def \ur {\hat \vr}
\def \Ru {R}
\def \Rsigma {\Delta \inc}
\def \Rmean {\langle \inc \rangle}
\def \Rmeani {\langle \inc_1 \rangle}
\def \Rmeanj {\langle \inc_2 \rangle}

\begin{document}	

   \title{Effect of the inclination in the passage through the 5/3 mean motion resonance between Ariel and Umbriel}
   
   \titlerunning{Effect of the inclination in the 5/3~MMR between Ariel and Umbriel}
   
   \authorrunning{S. R. A. Gomes \& A. C. M. Correia}

   \author{S\'ergio R. A. Gomes
          \inst{1}
          \and
          Alexandre C. M. Correia\inst{1,2}
          }

   \institute{CFisUC, Departamento de F\'isica, Universidade de Coimbra, 3004-516 Coimbra, Portugal
         \and
             IMCCE, UMR8028 CNRS, Observatoire de Paris, PSL Universit\'e, 77 Av. Denfert-Rochereau, 75014 Paris, France}

   \date{Received 7 February 2023/ Accepted 7 April 2023}

  \abstract{
The orbits of the main satellites of Uranus are expected to slowly drift away owing to tides raised in the planet. As a result, the 5/3 mean motion resonance between Ariel and Umbriel was likely encountered in the past. Previous studies have shown that, in order to prevent entrapment in this resonance, the eccentricities of the satellites must be larger than $\sim 0.01$ at the epoch, which is hard to explain. On the other hand, if the satellites experience some temporary capture and then escape, the inclinations rise to high values that are not observed today. We have revisited this problem both analytically and numerically focussing on the inclination, using a secular two-satellite model with circular orbits. We show that if the inclination of Umbriel was around $0.15\degree$ at the time of the 5/3 resonance encounter, capture can be avoided in about 60\% of the cases. Moreover, after the resonance crossing, the inclination of Umbriel drops to a mean value around $0.08\degree$, which is close to the presently observed one. The final inclination of Ariel is  distributed between $0.01\degree$ and $0.25\degree$ with a nearly equal probability, which includes the present mean value of $0.02\degree$.
}


   \maketitle

\section{Introduction}

The origin of the Uranian satellite system is not yet completely understood and it is still under debate \citep[eg.][]{Pollack_etal_1991, Szulagyi_etal_2018, Ishizawa_etal_2019, Inderbitzi_etal_2020, Ida_etal_2020, Rufu_Canup_2022}.
Despite the formation mechanism of the main satellites, their orbits  slowly drift away owing to tides raised in the planet \citep[eg.][]{Peale_1988, Tittemore_Wisdom_1988, Tittemore_Wisdom_1989, Tittemore_Wisdom_1990, Pollack_etal_1991, Cuk_etal_2020}.
Because the orbits do not all evolve at the same pace, 
they may have encountered several mean motion resonances (MMRs) in their way since their formation about $4.5$~Gyr ago. 
At present, there is no orbital commensurability between these satellites, but MMRs are often invoked to explain some anomalous observations, such as resurfacing events \citep[e.g][]{Dermott_etal_1988, Peale_1988, Tittemore_1990}, the current relatively large eccentricities ($\sim 0.001$) of all satellites \citep[e.g][]{Squyres_etal_1985, Smith_etal_1986, Peale_1988}, or the high inclination of Miranda ($\sim 4.3\degree$) \citep[e.g][]{Tittemore_Wisdom_1989, Tittemore_Wisdom_1990, Malhotra_Dermott_1990, Verheylewegen_etal_2013, Cuk_etal_2020}.

The latest low-order commensurability to have occurred was the 5/3~MMR between Ariel and Umbriel \citep[eg.][]{Peale_1988, Cuk_etal_2020}, which is consistent with recent geologic activity observed in Ariel \citep[eg.][]{Zahnle_etal_2003, Cartwright_etal_2020}.
The current free eccentricities still observed also support this possibility \citep[eg.][]{Jacobson_2014}.
A detailed study on the passage through this resonance was carried out by \citet{Tittemore_Wisdom_1988} using a secular resonant two-satellite planar model with small eccentricities.
It was shown that if the 5/3~MMR is approached with eccentricities of both satellites smaller than $\sim 0.01$, a long-term capture in resonance is certain. 
Therefore, in order to evade or skip this resonance, at least one of the eccentricities must have been close to 0.01 at the time.
However, \citet{Tittemore_Wisdom_1988} recognised that it is very difficult to justify such high initial eccentricity values.
Indeed, mutual perturbations between the largest Uranian satellites cannot explain such high values alone \citep[eg.][]{Dermott_Nicholson_1986, Laskar_1986}. 
Moreover, tidal friction within the satellites circularise the orbits in a short timescale \citep[eg.][]{Squyres_etal_1985}, and so any large eccentricity remnant left by a prior MMR crossing is expected to be quickly eroded. 

\citet{Cuk_etal_2020} revisited the passage through the 5/3~MMR using a $N$-body numerical integrator, which includes the five main satellites, non-planar orbits, and spin evolution.
They started their simulations with the current eccentricities and adopted nearly zero inclination for all satellites ($< 0.1\degree$). 
They confirm that low initial eccentricities translate into capture in resonance, which can nevertheless be broken after some time due to some chaotic excitation of the eccentricities.
They also observed that all five moons had their inclinations excited during the resonance entrapment.
The effect on the inclination is particularly significant in Miranda because it has the smallest mass; however, after leaving the resonance, the remaining four moons were all left with inclination values higher than the current ones.
In particular, Ariel and Umbriel acquired inclinations around $1\degree$, which cannot be damped to the current observed mean values of $\sim 0.02\degree$ and $0.08\degree$, respectively.
\citet{Cuk_etal_2020} then propose that the inclination of Umbriel can be lowered if its node is involved in a secular spin-orbit resonance with the spin of Oberon (not observed today). However,  even if this mechanism works, it would fail to damp the inclinations of the remaining satellites.

\citet{Cuk_etal_2020} suggest that the 5/3~MMR between Ariel and Umbriel can be responsible for the current high inclination of Miranda instead of the crossing of the 3/1~MMR between Miranda and Umbriel \citep{Tittemore_Wisdom_1989, Tittemore_Wisdom_1990}. 
However, the high inclination values also acquired by Ariel and Umbriel seem difficult to conciliate with the present observed system. 
A more plausible scenario is that Miranda's inclination is excited during the Miranda and Umbriel 3/1~MMR crossing, and then Ariel and Umbriel skip the 5/3~MMR without being captured to prevent any damage on the inclinations.
The problem with this scenario is that for initial near zero inclination values for both Ariel and Umbriel, at least one of these satellites must have an initial eccentricity larger than $0.01$ to avoid entrapment in the 5/3~MMR \citep{Tittemore_Wisdom_1988}, which is also challenging to explain.

Since the inclination appears to play an important role in the passage of the 5/3~MMR between Ariel and Umbriel, here we intend to look at this problem with a different perspective.
Instead of assuming a coplanar model with low initial eccentricities for Ariel and Umbriel as in \citet{Tittemore_Wisdom_1988}, we assume a circular model with low inclinations using a similar approach to \citet{Tittemore_Wisdom_1989} for the study of the 3/1~MMR between Miranda and Umbriel.
In Sect.~\ref{sec:Resonance_dynamics}, we introduce a secular resonant two-satellite circular model with low inclinations using complex Cartesian coordinates and describe all dynamical features present in the two degree-of-freedom phase space for different ratios of the semi-major axes.
In Sect.~\ref{sec:Tidal_evolution}, we add tidal effects to the equations of motion using a constant time-lag model and then compute the capture probabilities in the 5/3~MMR for different initial inclination values.
In Sect.~\ref{sec:Numerical_integration}, we perform a large number of numerical simulations to estimate the possible outcomes of the passage through the 5/3~MMR as a function of the initial inclinations.
Finally, in the last section, we summarise and discuss our results.

\section{Resonant secular dynamics}\label{sec:Resonance_dynamics}

In this section we study the conservative dynamics of a three-body system involved in a second order $(p+q)/p$~MMR, hence $q=2$.
For the 5/3~MMR, we additionally have $p=3$, but our model is valid for any $p$ value.
We further assume that the orbits are circular (zero eccentricities) and have low inclinations.

\subsection{Hamiltonian}

We consider an oblate central body of mass $m_0$ (Uranus) surrounded by two point-mass bodies $m_1$, $m_2 \ll m_0$ (satellites), where the subscript 1 refers to the inner orbit (Ariel) and the subscript 2 refers to the outer orbit (Umbriel). 
The potential energy of the system is given by \citep[eg.][]{Smart_1953}
\begin{equation}\label{eq:general_conservative_gravitational_potential}
    \begin{split}
    U=-\sum_{k=1}^2\frac{\mathcal{G}m_0m_k}{r_k}\left[1+ J_2\left(\frac{\Ru}{r_k}\right)^2P_2(\mathbf{\hat{r}}_k\cdot \vs)\right]
    -\frac{\mathcal{G}m_1m_2}{|\mathbf{r}_2-\mathbf{r}_1|} \ ,
    \end{split}
\end{equation}
where $\mathcal{G}$ is the gravitational constant; $J_2$, $\Ru$, and $\vs$ are the second order gravity field, the radius, and the spin unit vector of the central body, respectively; $\vr_k$ is the position vector of $m_k$ with respect to the centre of mass of $m_0$ (planetocentric coordinates); $r_k = |\vr_k|$ is the norm; $\ur_k = \vr_k / r_k $ is the unit vector; and $P_2(x)=(3x^2-1)/2$ is the Legendre polynomial of degree two. 
We neglected terms in $(\Ru/r_k)^3$ (quadrupolar approximation for the oblateness).
The Hamiltonian of the problem, $\Ham$, was then obtained by adding the orbital and rotational kinetic energies to Eq.\,(\ref{eq:general_conservative_gravitational_potential}) .

\subsubsection{Expansion in elliptical elements}

\label{sec:eee}

We can expand the Hamiltonian in elliptical elements. 
To the first order in the mass ratios, $m_k/m_0$, zeroth order in the eccentricities, and second order in the inclinations, $\inc_k$ (with respect to the equatorial plane of the central body), we have \citep[eg.][]{Murray_Dermott_1999}
\begin{equation}\label{genHam}
    \Ham = \Ham_K + \Ham_O + \Ham_S + \Ham_R + \Ham_F + \frac{\Theta^2}{2 C} \ ,
\end{equation}
where
\begin{equation}\label{Hkep}
    \Ham_K=-\sum_{k=1}^2\frac{\mathcal{G}m_0m_k}{2a_k}
\end{equation}
is the Keplerian part and $a_k$ are the semi-major axes,
\begin{equation}\label{Hobl}
    \Ham_O=-\sum_{k=1}^2\frac{\mathcal{G}m_0m_k}{2a_k}J_2\left(\frac{\Ru}{a_k}\right)^2\left(1-3\inc_k^2\sin^2{(\lambda_k-\Omega_k)}\right) 
\end{equation}
 is the contribution from the oblateness of the central body, $\lambda_k$ are the mean longitudes, $\Omega_k$ are the longitudes of the nodes,
\begin{equation}\label{Hsec}
    \begin{split}
        \Ham_S=-\frac{\mathcal{G}m_1m_2}{8a_2}\bigg[4 \, b_{\frac{1}{2}}^{(0)}(\alpha)- & \alpha \, b^{(1)}_{\frac{3}{2}}(\alpha) \, \bigg( \inc_1^2 +  \inc_2^2 \\
        & - 2 \inc_1 \inc_2 \cos{(\Omega_2-\Omega_1)}\bigg) \bigg] \,
    \end{split} 
\end{equation}
is the secular part, $b^{(j)}_s$ are Laplace coefficients, $\alpha=a_1/a_2$, and 
\begin{equation}\label{Hres}
\begin{split}
    \Ham_R= -\frac{\mathcal{G}m_1m_2}{8a_2} & \bigg[\inc_1^2\cos{((p+2)\lambdaj-p\lambdai-2\Omega_1)} \\
        & -2 \inc_1 \inc_2 \cos{((p+2)\lambdaj-p\lambdai-\Omega_2-\Omega_1)} \\
        & + \inc_2^2\cos{((p+2)\lambdaj-p\lambdai-2\Omega_2)} \bigg]  \alpha b^{(p+1)}_{\frac{3}{2}}(\alpha)
\end{split}
\end{equation}
is the contribution from the second order resonant terms ($q=2$).
The $\Ham_S$ and $\Ham_R$ terms arise solely from the direct part of the disturbing function (last term in Eq.\,(\ref{eq:general_conservative_gravitational_potential})), because in the circular approximation the indirect part does not have secular or second order resonant terms.
The term in $\Ham_F$ corresponds to the remaining terms of the disturbing function that depend on other combinations of the angles $\lambdai$, $\lambdaj$, $\Omega_1$, and $\Omega_2$ that do not appear in the expressions of $\Ham_S$ or $\Ham_R$.

Finally, for the last term in the Hamiltonian (Eq.\,(\ref{genHam})), corresponding to the rotational kinetic energy, $C$ is the principal moment of inertia of the central body and 
\begin{equation}\label{Lrot}
 \Theta = C \angrotU \ ,
\end{equation}
which corresponds to the rotational angular momentum, $\angrotU = \dot \theta$ is the angular velocity, and $\theta$ is the rotation angle.
In the conservative case, the rotational kinetic energy is constant and could be dropped from the Hamiltonian.
Nevertheless, when we include tidal dissipation (Sect.~\ref{sec:Tidal_evolution}), there are angular momentum exchanges between the spin and the orbits, and this term cannot be neglected.

\subsubsection{Action-angle resonant variables}

\label{sec:aarv}

We can rewrite the Hamiltonian (\ref{genHam}) using a set of canonical action-angle variables. 
For that purpose, we adopted Andoyer variables for the rotation ($\Theta, \theta$) and Poincar\'e variables for the orbits $(\Lambda_k,\lambda_k;\Psia_k,-\Omega_k)$ with
\begin{equation}\label{Lambdak}
\Lambda_k = \beta_k\sqrt{\mu_k a_k} \ ,
\end{equation}
\begin{equation}\label{Psik}
\Psia_k = \Lambda_k (1-\cos \inc_k) \approx \frac{1}{2} \Lambda_k \inc_k^2 \ ,
\end{equation}
 where $\beta_k=m_0m_k/(m_0+m_k)$ and $\mu_k=\mathcal{G}(m_0+m_k)$.
 Since we aim to study the dynamics of the system near the $(p+2)/p$~MMR, we introduce the near resonant angle
\begin{equation}\label{eq:resonace_argument}
    \sigma = \left(1+\frac{p}{2}\right)\lambda_2-\frac{p}{2}\lambda_1 \ ,
\end{equation}
which is present in all terms of the resonant Hamiltonian (Eq.\,(\ref{Hres})).
Each term corresponds to a resonant combination:
\begin{equation}\label{psiiang}
\psii = \sigma - \Omega_1 \ , \quad \psij = \sigma - \Omega_2 \ , \quad \psia_3 = \frac12 (\psii + \psij) \ .
\end{equation}
We further introduce the angles
\begin{equation}\label{addangles}
\gamma =\frac{p}{2} \left( \lambda_1 - \lambda_2 \right) =  \lambda_2 - \sigma  
\quad \mathrm{and} \quad  \vartheta = \theta - \sigma \ , 
\end{equation}
such that
\begin{equation}
\left[\begin{array}{c} 
\sigma \\  \gamma \\ \psii \\ \psij \\ \vartheta
\end{array}\right] 
\equiv {\cal S} \,
\left[\begin{array}{c} 
\lambda_1 \\ \lambda_2 \\ -\Omega_1 \\ -\Omega_2 \\ \theta
\end{array}\right] \ ,
\end{equation}
with
\begin{equation}
{\cal S} =  \left[\begin{array}{ccccc} 
-p/2 & 1 + p/2 & 0 & 0 & 0 \\
 p/2 & -p/2 & 0 & 0 & 0 \\
-p/2 & 1 + p/2 & 1 & 0 & 0 \\
-p/2 & 1 + p/2 & 0 & 1 & 0 \\
 p/2 & -1 - p/2 & 0 & 0 & 1
\end{array}\right]  \ ,
\end{equation}
which gives for the conjugated actions \citep[eg.,][]{Goldstein_1950}
\begin{equation}
\left[\begin{array}{c} 
\Sigma \\ \Gamma \\ \tilde \Psia_1 \\ \tilde \Psia_2 \\ \tilde \Theta
\end{array}\right] 
= ( {\cal S}^{-1} )^T \,
\left[\begin{array}{c} 
\Lambdai \\ \Lambdaj \\ \Psii \\ \Psij \\ \Theta
\end{array}\right] \ .
\end{equation}
The new set of canonical variables that uses the resonant angles is then given by
\begin{equation}
 \label{canonic_var}
\begin{array}{l l} 
\Sigma = \Lambdai + \Lambdaj - \Psii - \Psij + \Theta \ , & \sigma = \left(1+\frac{p}{2}\right)\lambda_2-\frac{p}{2}\lambda_1 \\
\Gamma = \left(1+\frac{2}{p}\right)\Lambdai+\Lambdaj \ , & \gamma =  \lambda_2 - \sigma  \\
\tilde \Psia_1 = \Psii  \ , & \psii = \sigma - \Omega_1  \\
\tilde \Psia_2 = \Psij  \ , & \psij = \sigma - \Omega_2  \\
 \tilde \Theta = \Theta \ , & \vartheta = \theta - \sigma 
\end{array} \ .
\end{equation}

\begin{table*}
\caption{Present physical and orbital properties of the Uranian system \citep{Thomas_1988, Jacobson_2014}.}
\label{table:physical_orbital_parameters}      
\centering          
\begin{tabular}{c c c c c c c}     
	\hline\hline       
    &$m$ (M\textsubscript{\(\odot\)}$\times10^{-10}$)	& $\Ru$ (km) & $\langle T_{\rm rot}\rangle$ (day)& $J_2 (\times10^{-3})$ & $C/(m_0R^2)$ &\\
    \hline
    Uranus & 436562.8821 & 25559. & 0.7183 & 3.5107 & 0.2296\\
	\hline\hline       
	Satellite 	& $m$ (M\textsubscript{\(\odot\)}$\times10^{-10}$)	& $\Ru$ (km) & $\langle T_{\rm orb}\rangle$ (day) & $\langle a \rangle$ ($\Ru$) & $\langle e \rangle$ ($\times 10^{-3}$) & $\langle \inc \rangle$ ($\degree$)\\
	\hline                    
	Ariel 	& 6.291561 	& 578.9  & 2.479971 	& 7.468180 	& 1.22 & 0.0167\\
	Umbriel & 6.412118 	& 584.7  & 4.133904 	& 10.403550	& 3.94 & 0.0796\\
    \hline
\end{tabular}
\end{table*}

\subsubsection{Conserved quantities and average}

\label{cqaa}

We can rewrite the Hamiltonian (\ref{genHam}) using the resonant canonical variables (Eq.\,(\ref{canonic_var})).
For the actions, we can replace the semi-major axes and the inclinations using relations (\ref{Lambdak}) and (\ref{Psik})
\begin{equation}\label{smak}
 a_k = \frac{\Lambda_k^2}{\beta_k^2 \mu_k} \ ,
\end{equation}
\begin{equation}\label{inckL}
 \inc_k \approx \sqrt{\frac{2 \Psia_k}{\Lambda_k}}  \ .
\end{equation}
It is important to note, however, that the $\Lambda_k$ are no longer actions of the resonant variables, and they must be obtained as (Eq.\,(\ref{canonic_var}))
\begin{equation}\label{Lb1}
\Lambdai  = \Gamma_1 - \frac{p}{2} \left( \Psii + \Psij \right)  \ ,
\end{equation}
\begin{equation}\label{Lb2}
\Lambdaj = \Gamma_2 + \left(1+\frac{p}{2}\right) \left( \Psii + \Psij \right)  \ ,
\end{equation}
with
\begin{equation}\label{Gb1}
\Gamma_1  = \frac{p}{2} \Gamma \left( 1 - \Delta \right) \ ,
\end{equation}
\begin{equation}\label{Gb2}
\Gamma_2 = -\frac{p}{2} \Gamma \left(1 -\left(1+\tfrac{2}{p}\right) \Delta \right)  \ ,
\end{equation}
and
\begin{equation}\label{DeltaRef}
\Delta = \left(\Sigma - \Theta\right) / \Gamma \ .
\end{equation}
In the approximation of low inclinations, 
$ \Psia_k \ll \Lambda_k$ (Eq.\,(\ref{inckL})),
and so we also have $ \Psia_k \ll \Gamma_k$, allowing us to write
\begin{equation}\label{Lba}
\Lambdai^\alpha \approx \Gamma_1^\alpha \left[1 - \alpha \frac{p}{2} \frac{\Psii + \Psij }{\Gamma_1} \right] \ ,
\end{equation}
\begin{equation}\label{Lba}
\Lambdaj^\alpha \approx \Gamma_2^\alpha \left[1 + \alpha \left(1+\frac{p}{2}\right) \frac{\Psii + \Psij}{\Gamma_2} \right] \ ,
\end{equation}
and
\begin{equation}\label{inck}
 \inc_k \approx \sqrt{\frac{2 \Psia_k}{\Gamma_k}}  \ .
\end{equation}

The angle $\theta$ does not appear in the expression of the Hamiltonian (Eq.\,(\ref{genHam})), and so $\vartheta$ does not appear either (Eq.\,(\ref{canonic_var})).
As a consequence, the conjugated action, $\Theta$, is a constant of motion.
According to the d'Alembert rule (conservation of the angular momentum), the remaining angles must be combined as
\begin{equation}\label{dAlembert1}
\cos{\big(k_1 \lambdai + k_2 \lambdaj + k_3 \Omega_1 + k_4 \Omega_2\big)} \ , \quad k_i \in \mathbb{Z} \ ,
\end{equation}
where $k_1+k_2+k_3+k_4=0$, or, using the resonant angles (Eq.\,(\ref{canonic_var})),
\begin{equation}\label{dAlembert2}
 \cos{\left( \Big(k_1+\frac{2}{p} k_1 + k_2 \Big) \, \gamma - k_3 \psii + \Big(k_1+k_2+k_3\Big) \, \psij\right)} \ .
\end{equation}
Thus, the angle $\sigma$ also does not appear in the Hamiltonian, and its conjugated action, $\Sigma$, is a constant of motion.
We note that,
\begin{equation}\label{SigmaTOT}
    \begin{split}
\Sigma & = \left(\Lambdai - \Psii\right) + \left(\Lambdaj - \Psij\right) + \Theta \\
& = \beta_1 \sqrt{\mu_1 a_1} \cos \inc_1 + \beta_2 \sqrt{\mu_2 a_2} \cos \inc_2 + C \angrotU
    \end{split} 
\end{equation}
does indeed correspond to the projection of the total angular momentum of the system along the direction of $\vs$, which must be conserved.

Near the MMR, the two resonant angles $\psii$ and $\psij$ present a slower variation than the angle $\gamma$, that is, $\dot \psii, \dot \psij \ll \dot \gamma$.
Therefore, we can construct the resonant secular normal form of the Hamiltonian (to the first order in $m_k/m_0$) by averaging over $\gamma$:
\begin{equation}\label{averHam}
\avHam = \langle \Ham \rangle_\gamma = \frac{1}{2 p \pi} \int_0^{2 p \pi} \Ham \, d \gamma \ . 
\end{equation}
As a result, $\langle \Ham_F \rangle_\gamma = 0$, and since $\gamma$ no longer appears in the expression of the averaged Hamiltonian, the conjugated variable, $\Gamma$ (Eq.\,(\ref{canonic_var})), also becomes a constant of motion.
We thus reduced a problem with five degrees of freedom initially to a problem with only two degrees of freedom, ($\Psii, \psii$) and ($\Psij, \psij$), and three parameters, $\Sigma$, $\Gamma$, and $\Theta$.
The auxiliary quantities $\Gamma_1$ (Eq.\,(\ref{Gb1})) and $\Gamma_2$ (Eq.\,(\ref{Gb2})) are also constant.

The resonant secular Hamiltonian then finally reads as follows:
\begin{equation}
\label{resHam}
    \begin{split}
       \avHam &=(\Ka+\Sa)(\Psii+\Psij)+\Kb(\Psii+\Psij)^2\\
        &+(\Oc+\Se)\Psii+(\Od+\Sf)\Psij\\
        &+\Sg\sqrt{\Psii}\sqrt{\Psij}\cos{(\psii-\psij)}\\
        &+\Rd\Psii\cos{(2\psii)}+\Re\Psij\cos{(2\psij)}\\
        &+\Rf\sqrt{\Psii}\sqrt{\Psij}\cos{(\psii+\psij)} \ ,
    \end{split} 
\end{equation}
where $\mathcal{K}$ stands for the Keplerian coefficients (Eq.\,(\ref{Hkep})), $\mathcal{O}$ for the oblateness coefficients (Eq.\,(\ref{Hobl})), $\mathcal{S}$ for secular the coefficients (Eq.\,(\ref{Hsec})), and $\mathcal{R}$ for the resonant coefficients (Eq.\,(\ref{Hres})).
The Keplerian part needs to be expanded to the second order in $\Psia_k$ (fourth order in the inclinations), because $\Kb$ is much larger than the remaining coefficients.
The explicit expression of all these coefficients is given in appendix~\ref{sec:Conservative_Hamiltonian_terms}.

\subsubsection{Complex rectangular coordinates}\label{sec:rectangular_complex_coordinates}

\label{sec:crc}

The equations of motion expressed in the variables $(\Psia_k, \psia_k)$ may experience some singularities when $\Psia_k=0$ (because of the terms in $\Sg$ and $\Rf$).
To remove this problem, we can perform a second canonical change of variables to rectangular coordinates 
$(\Psia_k, \psia_k) \rightarrow (\yk, \ii \ybk)$, where
\begin{equation}\label{eq:complex_cartesian_coordinates}
    \yk = \sqrt{\Psia_k}\ep^{\ii\psia_k} \ ,
\end{equation}
and $\ybk$ is the complex conjugate of $\yk$.
From Eq.\,(\ref{inck}), we have 
\begin{equation}\label{yinc}
\yk \approx \inc_k \sqrt{\frac{\Gamma_k}{2}} \ep^{\ii\psia_k}  \ ,
\end{equation}
and so these variables are proportional to the inclinations.
The resonant secular Hamiltonian (Eq.\,(\ref{resHam})) now reads as follows: 
\begin{equation}
\label{cartHam}
    \begin{split}
       \avHam&=\left(\Ka+\Sa\right)(\yi\ybi+\yj\ybj)+\Kb(\yi\ybi+\yj\ybj)^2\\
        &+\left(\Oc+\Se\right) \yi\ybi+\left(\Od+\Sf\right) \yj\ybj+\frac{\Sg}{2}(\yi\ybj+\ybi\yj)\\
        &+\frac{\Rd}{2}(\yi^2+\ybi^2)+\frac{\Re}{2}(\yj^2+\ybj^2)+\frac{\Rf}{2}(\yi\yj+\ybi\ybj) \ .
    \end{split}
\end{equation}
The conservative equations of motion are simply obtained as
\begin{equation}\label{eq:gen_equations}
\frac{d \yk}{d t} = \ii \frac{\partial \avHam}{\partial \ybk} \ ,
\end{equation}
yielding
\begin{equation}\label{eq:conservative_motion_equations1}
    \begin{split}
    \dot{y}_{1}=\ii \, &\Bigg[\left(\Ka+\Sa\right)\yi+2\Kb\left(\yi\ybi+\yj\ybj\right)\yi\\
    &+\left(\Oc+\Se\right)\yi+\frac{\Sg}{2}\yj+\Rd\ybi+\frac{\Rf}{2}\ybj\Bigg] 
    \end{split}
\end{equation}   
and 
\begin{equation}\label{eq:conservative_motion_equations2}
    \begin{split}
    \dot{y}_{2}=\ii \, &\Bigg[\left(\Ka+\Sa\right)\yj+2\Kb\left(\yi\ybi+\yj\ybj\right)\yj\\
    &+\left(\Od+\Sf\right)\yj+\frac{\Sg}{2}\yi+\Re\ybj+\frac{\Rf}{2}\ybi \Bigg] \ .
    \end{split}
\end{equation}

\subsection{Dynamical evolution}

The values of $m_k$, $J_2$, and $\Ru$ are relatively well determined for the Uranian system (Table \ref{table:physical_orbital_parameters}).
Therefore, to compute the coefficients $\mathcal{K}$, $\mathcal{O}$, $\mathcal{S}$, and $\mathcal{R}$ appearing in the Hamiltonian (\ref{cartHam}), we only need to know the values of the parameters $\Gamma_1$ and $\Gamma_2$ (see Appendix~\ref{sec:Conservative_Hamiltonian_terms}), which in turn depend on the parameters $\Gamma$ and $\Delta$ (Eqs.\,(\ref{Gb1}) and (\ref{Gb2})).

If we neglect the $\mathcal{O}$ coefficients (oblateness coefficients), it is possible to eliminate the dependence in $\Gamma$, because we get $\avHam \propto \Gamma^{-2}$ \citep[see appendix~\ref{sec:Conservative_Hamiltonian_terms} and][Sect.~2.1.4]{Delisle_etal_2012}.
Although here we cannot use this simplification, the conservative dynamics is still not very sensitive to the $\Gamma$ parameter \citep[eg.][]{Tittemore_Wisdom_1988, Tittemore_Wisdom_1989}, and so we fixed it at the reference value\footnote{The value of $\Gamma$ was obtained by reversing the orbital tidal evolution of the system (Eq.\,(\ref{eq:semi_major_axix_tidal_evolution})) until the nominal resonance was achieved (Eq.\,(\ref{nominal:sma_back})) and starting with the present semi-major axes (see Sect.~\ref{subsec:resonratios}).} 
\begin{equation}\label{Gammafix}
\Gamma = \num{2.6684e-12} \ \mathrm{M_\odot \, au^2 \, yr^{-1}} \ .
\end{equation}

The dynamics of the 5/3~MMR essentially depends on the $\Delta$ parameter (Eq.\,(\ref{DeltaRef})), which measures the proximity to the resonance.
Following \citet{Delisle_etal_2012}, we write
\begin{equation}\label{delta:equation}
\delta = \frac{\Delta}{\Delta_r} - 1 \ ,
\end{equation}
where $\Delta_r$ is the value of $\Delta$ at the planar ($\inc_k=0$) nominal resonance, that is (Eq.\,(\ref{DeltaRef}) and (\ref{SigmaTOT})), 
\begin{equation}\label{Deltari}
\Delta_r = \left(\Lambda_{1,r} + \Lambda_{2,r}\right) / \Gamma_r \ ,
\end{equation}
when $n_1/n_2=5/3$, where $n_k$ is the mean motion of the satellite with mass $m_k$.
At the nominal resonance, using Kepler's third law, we have the following relation for the $\Lambda_{k,r}$:
\begin{equation}\label{keplerreson}
\Lambda_{2,r} = \epsilon \left(\tfrac{5}{3}\right)^{1/3} \Lambda_{1,r} \ , \quad \mathrm{with} \quad \epsilon = \frac{\beta_2 \mu_2^{2/3}}{\beta_1 \mu_1^{2/3}} \approx \frac{m_2}{m_1} \ .
\end{equation}
From the expression of $\Gamma_r$ (Eq.\,(\ref{canonic_var})), we additionally have 
\begin{equation}\label{GammaRnom}
\Lambda_{2,r} = \Gamma_r - \frac53 \Lambda_{1,r} \ ,
\end{equation}
which can be combined with expression (\ref{keplerreson}) to give
\begin{equation}
 \Lambda_{1,r}  = \left( \tfrac53 + \epsilon \left(\tfrac53\right)^{1/3} \right)^{-1} \Gamma_r
\end{equation}\begin{equation}
\Lambda_{2,r} = \left( 1+ \epsilon^{-1} \left(\tfrac53\right)^{2/3} \right)^{-1} \Gamma_r  \ ,
\end{equation}
and finally (Eq.\,(\ref{Deltari}))
\begin{equation}\label{Deltarf}
\Delta_r = \left( 1 + \epsilon \left(\tfrac53\right)^{1/3} \right) \left( \tfrac53 + \epsilon \left(\tfrac53\right)^{1/3} \right)^{-1}   \ .
\end{equation}

\subsubsection{Equilibrium points}\label{subsec:equilibrium_points}

The equilibrium points correspond to stationary solutions of the Hamiltonian. 
They can be found by solving 
\begin{equation}\label{eq:equilibrium_points_condition_cartesian}
    \frac{\partial\avHam}{\partial\yi}=0 \quad \mathrm{and} \quad \frac{\partial\avHam}{\partial\yj}=0 \ ,
\end{equation}
that is, finding the roots of Eqs.~(\ref{eq:conservative_motion_equations1}) and (\ref{eq:conservative_motion_equations2}).
Splitting these equations into their real and imaginary parts, $\yk = y_{k,r} + \ii y_{k,i}$, we get

\begin{equation}\label{eq:y2r}
    y_{2,r} = - 2 \frac{\Ka+2\Kb\left(\Psii+\Psij\right)
    +\Oc+\Sa+\Se +\Rd}{\Sg + \Rf} y_{1,r}  \ ,
\end{equation}    
\begin{equation}\label{eq:y1r}
    y_{1,r}  = -2 \frac{\Ka+2\Kb\left(\Psii+\Psij\right)+\Od+\Sa+\Sf +\Re}{\Sg + \Rf} y_{2,r} \ ,
\end{equation}
\begin{equation}\label{eq:y2i}
    y_{2,i} = -2 \frac{\Ka+2\Kb\left(\Psii+\Psij\right)
    +\Oc+\Sa+\Se -\Rd}{\Sg-\Rf} y_{1,i} \ ,
\end{equation}    
\begin{equation}\label{eq:y1i}
    y_{1,i} = -2 \frac{\Ka+2\Kb\left(\Psii+\Psij\right)+\Od+\Sa
    +\Sf -\Re}{\Sg-\Rf} y_{2,i} \ .
\end{equation}

Replacing expressions (\ref{eq:y1r}) and (\ref{eq:y1i}) into expressions (\ref{eq:y2r}) and (\ref{eq:y2i}), respectively, yields that equilibria arise for
\begin{equation}\label{solzero}
y_{1,r}=y_{2,r}=0 \quad \mathrm{or} \quad y_{1,i}=y_{2,i}=0 \ .
\end{equation}
A more deep analysis shows that stable equilibria can only occur when the real roots are null.
We then focussed on these roots to determine the exact position of the stable equilibria.
Since $y_{1,r}=y_{2,r}=0$, we have (Eq.\,(\ref{eq:complex_cartesian_coordinates}))
\begin{equation}\label{imag_roots}
y_{1,i} = \pm \sqrt{\Psii} \quad \mathrm{and} \quad y_{2,i}= \pm \sqrt{\Psij} \ ,
\end{equation}
which we replaced in expressions (\ref{eq:y2i}) and (\ref{eq:y1i}) to determine the possible values of $\Psii$ and $\Psij$, \begin{equation}\label{root0}
\Psii=\Psij=0 
\end{equation}
or
\begin{equation}\label{root1}
\Psii = \frac{(\Rf-\Sg)\varepsilon_{\pm}-2(\Ka+\Oc+\Sa+\Se-\Rd)}{4 \Kb (1+\varepsilon^2_{\pm})} \ ,
\end{equation}
\begin{equation}\label{root2}
\Psij = \varepsilon_{\pm}^2 \, \Psii \ ,
\end{equation}
with 
\begin{equation}
\begin{split}
    \varepsilon_{\pm}=&\frac{\Oc - \Od - \Rd + \Re + \Se - \Sf}{\Rf - \Sg}\\
   &\pm\frac{ \sqrt{(\Oc - \Od - \Rd + \Re + \Se - \Sf)^2 + (\Rf - \Sg)^2}}{\Rf - \Sg} \ .
   \end{split}
\end{equation}

The equilibrium point at $\Psii=\Psij=0$ is always present, although it can be stable or unstable.
The remaining equilibria only exist for some $\delta$ values.
In Fig.~\ref{Fig:equilibrium_points}, we show the evolution of the equilibrium points as a function of $\delta$. We rescaled the $y_{k,i}$ by $\sqrt{\Gamma_k/2}$, such that we can translate the different equilibria in terms of inclinations (Eq.\,({\ref{yinc})).

For positive $\delta$ values far from zero, there is only one equilibrium point at $\inc_k=0$, which is stable (in blue).
For $\delta = \num{2.068e-6}$, there is a first bifurcation in the equilibria: two new stable equilibrium points appear at non-zero inclination (in green), while the point at $\inc_k=0$ becomes unstable.
For $\delta = \num{-2.540e-7}$, which is very close to the resonance nominal value $\delta=0$ (Eq.\,(\ref{Deltari})), a second bifurcation arises: two additional unstable equilibrium points appear at a non-zero inclination (in red), while the point at $\inc_k=0$ becomes stable again.

\begin{figure}
   \centering
   \includegraphics[width=\linewidth]{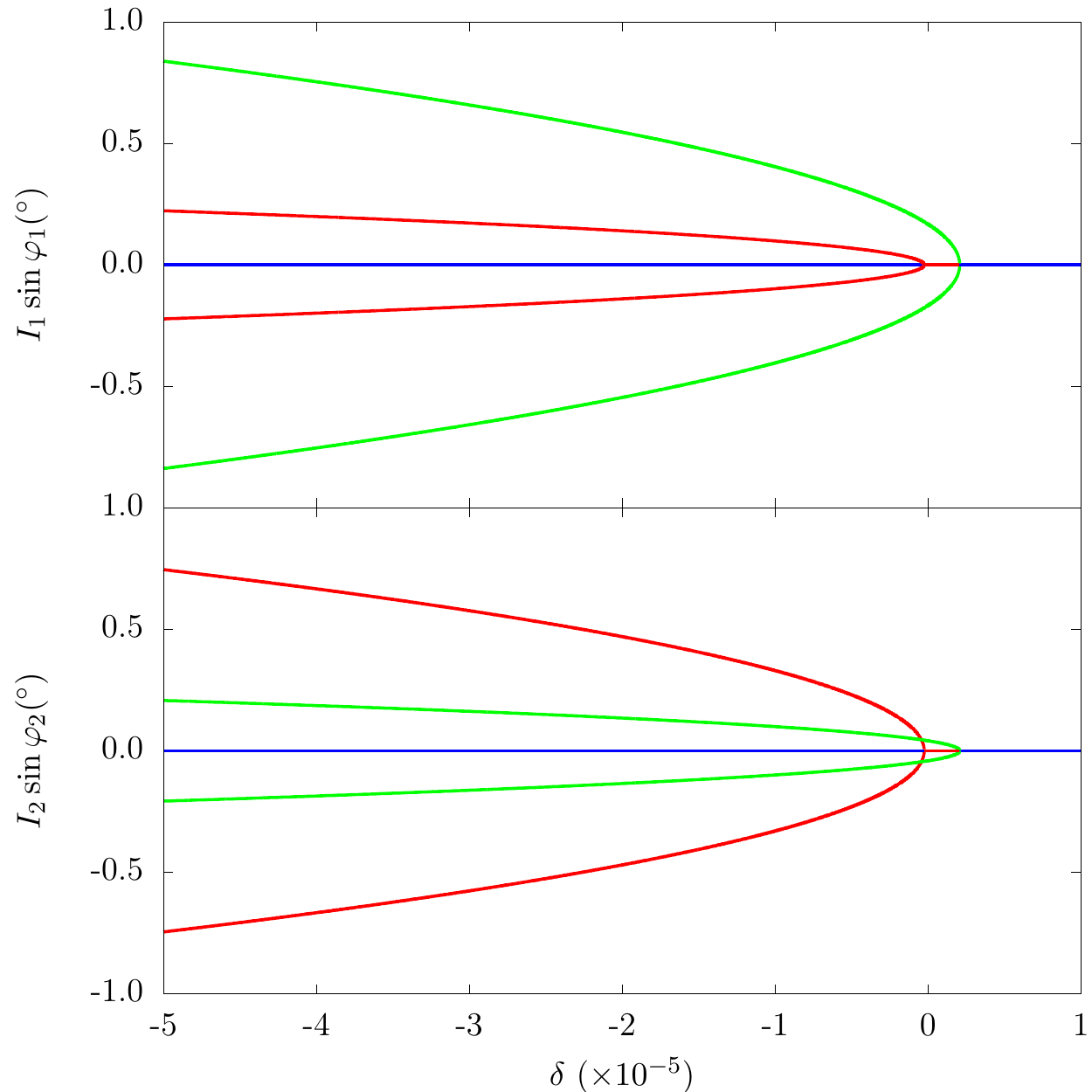}
      \caption{Evolution of the equilibrium points as a function of $\delta$. The green lines represent stable points inside the resonance (in a libration region), the red lines represent hyperbolic points (unstable), and the blue lines represent stable fixed points (in a circulation region).}
         \label{Fig:equilibrium_points}
   \end{figure}

\subsubsection{Energy levels}

For a better understanding of the dynamics, we can look at the energy levels of the resonant secular Hamiltonian (\ref{cartHam}) for different values of $\delta$ (Eq.\,(\ref{Deltarf})).
Since our problem has two degrees of freedom, and thus four dimensions, we need to plot these levels on section planes.
We chose the plane ($y_{1,i}, y_{2,i}$) with $y_{1,r}=y_{2,r}=0$, in order for stable equilibria to become visible (Eq.\,(\ref{imag_roots})).
In Fig.~\ref{fig:level_curves}, we show the energy levels for three values of $\delta$, which are representative of the three different equilibrium possibilities appearing in Fig.~\ref{Fig:equilibrium_points}.
Again, we rescaled $y_{k,i}$ by $\sqrt{\Gamma_k/2}$, and so we actually show the energy levels in the plane ($\inc_1 \sin \psii, \inc_2 \sin \psij $) with $\cos \psii = \cos \psij = 0$.

For $\delta = \num{5e-6} > 0 $ (Fig.~\ref{fig:level_curves}a), there is a single equilibrium point $(y_{1,i}=0, y_{2,i}=0)$ at the centre (in blue). 
It corresponds to a fixed point of the Hamiltonian (\ref{cartHam}), which is surrounded by a circulating region.
Therefore, in this case (and for higher $\delta$ values), all trajectories are outside the 5/3~MMR.

For $\delta=0$ (Fig.~\ref{fig:level_curves}b), the system is at the nominal resonance (Eq.\,(\ref{delta:equation})).
In this case, the equilibrium point at the centre  $(y_{1,i}=0, y_{2,i}=0)$ is still present (in red), but it is now unstable. 
Indeed, there is a separatrix in a tilted `8' shape emerging from this point that surrounds two additional stable equilibrium points (in green).
Trajectories inside the separatrix that encircle the stable points are in libration and correspond to orbits inside the 5/3~MMR.
Trajectories outside the separatrix are in circulation.

Finally, for $\delta=\num{-2e-6} < 0$ (Fig.~\ref{fig:level_curves}c), several equilibria exist.
There are two hyperbolic points (in red) from which a separatrix with two `banana' shapes emerges.
This separatrix delimits the phase space in libration and circulation regions.
There are two stable points (in green), one inside each banana island.
Trajectories that move around these points are in libration and correspond to orbits inside the 5/3~MMR.
The point $(y_{1,i}=0, y_{2,i}=0)$ at the centre (in blue) is again stable and inside a small circulation region.
Trajectories outside the separatrix are also in circulation.
This kind of phase space persists for smaller $\delta$ values, but the central circulation region becomes larger, while the resonant islands become thinner.

\begin{figure}
    \centering
    \includegraphics[width=\linewidth]{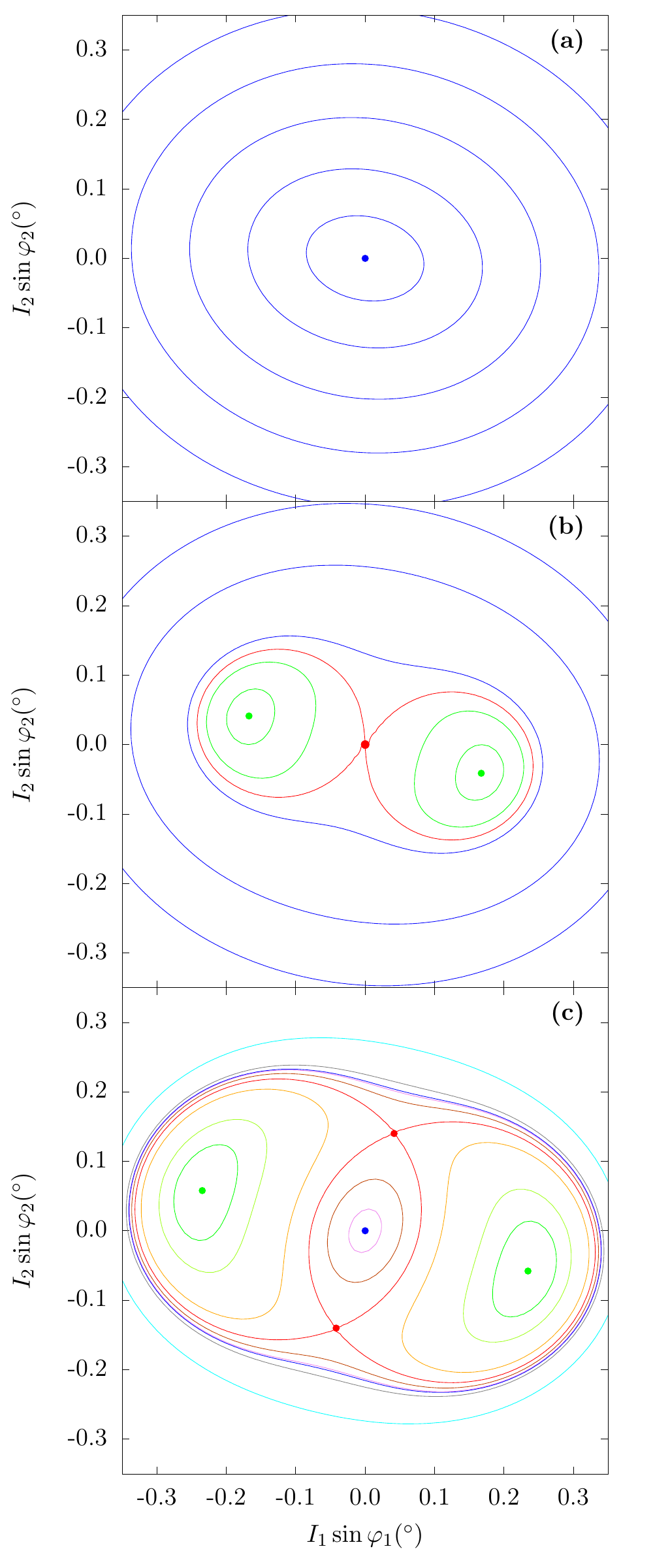}
    \caption{Energy level curves in the plane ($\inc_1 \sin \psii, \inc_2 \sin \psij$) with $\cos \psii = \cos \psij =0$, for $\delta=\num{5e-6}$ (top), $\delta=0$ (middle), and $\delta=\num{-2e-6}$ (bottom). Stable equilibria are coloured in green (resonance) and blue, while unstable equilibria are coloured in red, as well as the level curves that correspond to the separatrix.}
    \label{fig:level_curves}
\end{figure}

\subsubsection{Surface sections}
\label{sec:poincare}

The energy levels from previous the section allowed us to identify the different regions of the phase space, but a priori they do not correspond to trajectories followed by the system.
Indeed, since our problem has four dimensions, they show the trajectories when they cross the section plane with $y_{1,r}=y_{2,r}=0$, which only remain constant for the equilibrium points.
An alternative projection consists of fixing only $y_{1,r}=0$ (or $y_{2,r}=0$) together with a constant energy, that is, to draw Poincar\'e surface sections.
This projection is less restrictive, and therefore allows us to distinguish between periodic (fixed points), quasi-periodic (closed curves), and chaotic trajectories.
For that purpose, we used the modified H\'enon method \citep{Henon_Heiles_1964, Palaniyandi_2009}.

In Fig.~\ref{fig:level_curves}, we observe that the more diverse dynamics occurs for $\delta=-2 \times 10^{-6}$.
We thus adopted this value to draw the surface sections.
Since the Hamiltonian (\ref{cartHam}) is a four-degree function of $\yk$, the intersection of the constant energy manifold by a plane may have up to four roots (families).
Each family corresponds to a different dynamical behaviour, and so we must plot one of them at a time.
However, the families are symmetric and actually we only need to show two of them.
We chose to represent the families with the positive roots (that we dub 1 and 2).

In Fig.~\ref{fig:Poincare_surface_ariel}, we show a set of surface sections for the motion of Ariel in the plane ($y_{1,i}, y_{1,r}$) with $y_{2,r}=0$. 
We rescaled $\yk$ again by $\sqrt{\Gamma_k/2}$, and so we actually show the surface sections in the plane ($\inc_1 \sin \psii, \inc_1 \cos \psii $) with $\cos \psij = 0$.
Each panel corresponds to a different energy value, which coincide with the energy levels that are shown in Fig.~\ref{fig:level_curves}c (we adopted the same colour code).
The lowest energies occur in the circulation regions, $\avHam<\HamS$, the separatrix corresponds to a transition level, $\avHam=\HamS$, while the largest energies occur in the libration region, $\avHam>\HamS$.
The inner circulation region is delimited by the levels $0<\avHam<\HamS$, where $\avHam=0$ corresponds to the energy of the equilibrium point with $y_1=y_2=0$ (blue point in Fig.~\ref{fig:level_curves}c).
For this energy range, there are four families, while for the remaining energies only two families exist.

\begin{figure*}
    \centering
    \includegraphics[width=0.90\textwidth]{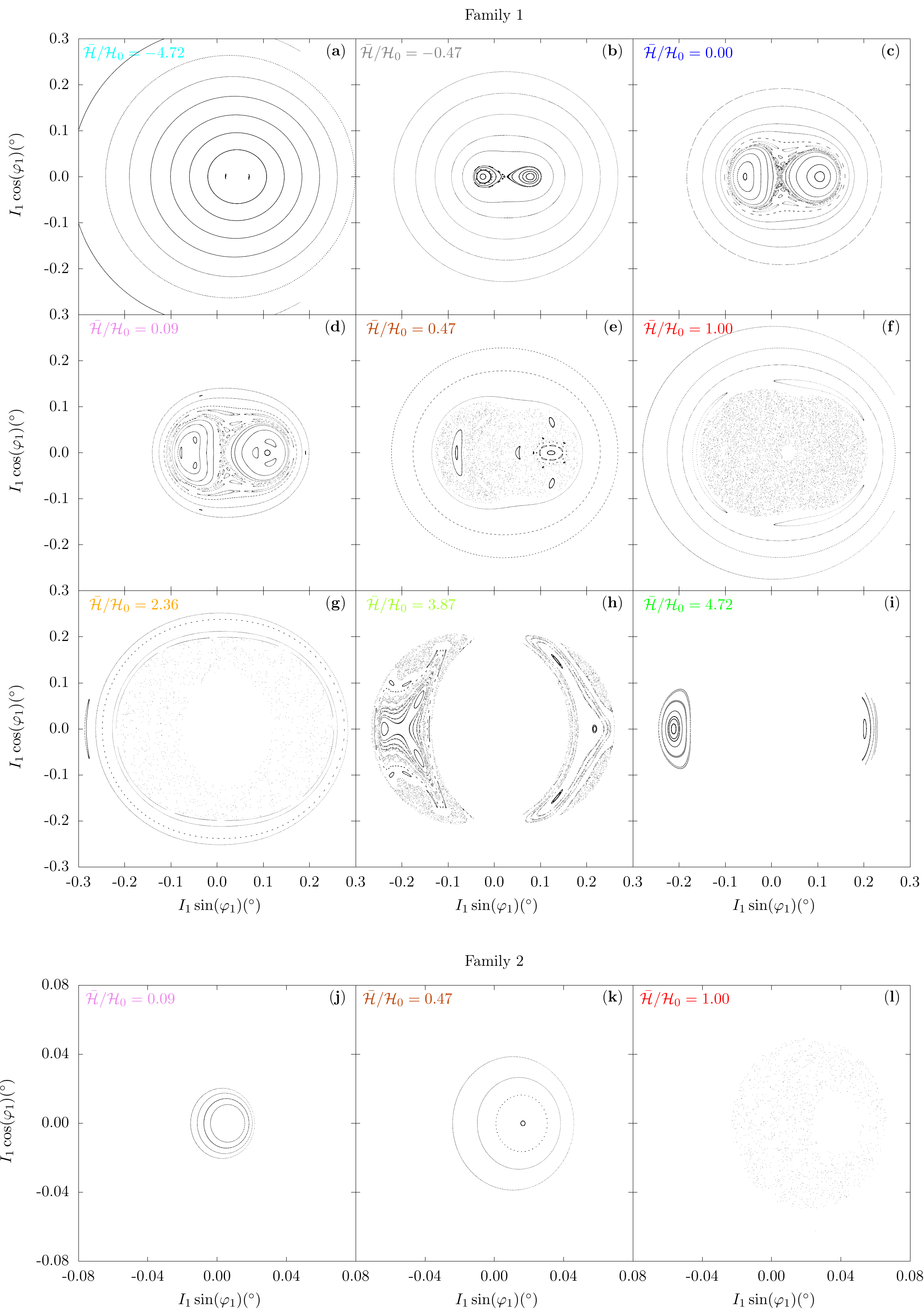}
    \caption{Poincar\'e surfaces of section for Ariel in the plane ($\inc_1 \sin \psii, \inc_1 \cos \psii $) with $\cos \psij=0$ and $\delta=\num{-2e-6}$. Each panel was obtained with a different energy value, corresponding to the energy levels shown in Fig.~\ref{fig:level_curves}c (we adopted the same colour code), and $\HamS=\num{1.06e-19} \ \mathrm{M_\odot \, au^2 \, yr^{-2}}$.}
    \label{fig:Poincare_surface_ariel}
\end{figure*}

For $\avHam \ll 0$ (Fig.~\ref{fig:Poincare_surface_ariel}a), only family 1 exists, and we observe that the system is always quasi-periodic, corresponding to trajectories in the outer circulation region.
As the energy increases, two islands appear, corresponding to trajectories that are in the libration region (in resonance).
Initially, the motion in these new regions is also quasi-periodic.
However, as the energy approaches the threshold $\avHam = 0$ (Fig.~\ref{fig:Poincare_surface_ariel}b,c), some chaotic regions appear in the transition between the circulation and libration regions.
For  $0<\avHam<\HamS$ (family 1), the chaotic regions increase, while the resonant islands shrink (Fig.~\ref{fig:Poincare_surface_ariel}d,e), until they completely disappear for $\avHam=\HamS$ (Fig.~\ref{fig:Poincare_surface_ariel}f).
For this specific energy range, we also needed to plot family 2.
Close to $\avHam = 0$, we observe quasi-periodic motion in the inner circulation region, but, as we approach $\avHam=\HamS$, this area is also completely replaced by a chaotic region (Fig.~\ref{fig:Poincare_surface_ariel}j,k,l).
Finally, for $\avHam>\HamS$, we observe that the chaotic region progressively vanishes and it is replaced by quasi-periodic motion in the libration region (Fig.~\ref{fig:Poincare_surface_ariel}g,h,i).
In this energy range, we only have family 1 and trajectories in the outer circulation region also do not exist.
Moreover, there is also a forbidden region at the centre of each panel that grows with the energy value while the libration areas shrink.

From the analysis of the surface sections, we conclude that the dynamics of the 5/3~MMR between Ariel and Umbriel is very rich and depends on the energy of the system.
In fact, the energy depends on the value of the inclinations (Eq.\,(\ref{cartHam})), given by the variables $y_1$ and $y_2$ (Eq.\,(\ref{yinc})).
Therefore, the value of the inclinations of Ariel and Umbriel when the system encounters the resonance can trigger completely different behaviours.
For $\avHam < 0$, the motion is quasi-periodic, either in circulation or libration.
Near the separatrix, $\avHam \sim \HamS$, the motion is mainly chaotic.
Finally, for $\avHam \gg \HamS$, the motion is again quasi-periodic, but only possible in libration with a small amplitude around the high inclination stable equilibrium points (Fig.~\ref{Fig:equilibrium_points}).

\section{Tidal evolution}\label{sec:Tidal_evolution}

The resonant dynamics presented in the previous section is conservative and thus the average semi-major axes remain constant.
However, the orbits of the Uranian satellites are expected to evolve because of tidal interactions.
The tidal contributions to the orbital and spin evolution can be obtained by considering an additional tidal potential \citep[eg.][]{Darwin_1880, Kaula_1964}.

\subsection{Tidal potential energy}

Tides arise from differential and inelastic deformations of an extended body (e.g. Uranus) owing to the gravitational effect of a perturber (e.g. Ariel or Umbriel).
The resulting distortion gives rise to a tidal bulge, which modifies the gravitational potential of the extended body. 
The dissipation of the mechanical energy of tides inside the body introduces a time delay, $\dtU$, between the initial perturbation and the maximal deformation. 
As the perturber interacts with the additional potential field, the amount of tidal potential energy is given by 
\begin{equation}\label{eq:tidal_potential}
    U_k=-k_2 \frac{\mathcal{G}m_{k}^2}{\Ru} \left(\frac{\Ru}{r_k}\right)^3 \left(\frac{\Ru}{r_k'}\right)^3 P_2 (\ur_k \cdot \ur_k')  \ ,
\end{equation}
where $k_2$ is the elastic second Love number for potential of the body, $\vr_k = \vr_k(t)$ is the position of the pertuber at a time $t$, and $\vr_k' = \vr_k(t-\dtU)$ is its position when it exerts the perturbation.
In this work we solely consider the deformations raised on Uranus by its satellites, since we assumed the satellites as point masses (Sect.~\ref{sec:Resonance_dynamics}).
This choice is fully justified because for circular orbits and synchronous satellites the tides raised by the planet on its satellites can be neglected \citep[eg.][]{Correia_2009}.

Although tidal effects do not preserve the mechanical energy, it is possible to extend the Hamiltonian formalism from Sect.~\ref{sec:Resonance_dynamics} by considering the primed quantities, $\vr_k'$, as parameters \citep{Mignard_1979}.
The tidal Hamiltonian then reads (Eqs.\,(\ref{genHam}) and (\ref{eq:tidal_potential})) as follows:
\begin{equation}
\label{tidalH}
    \Ham_t=\Ham + U_1 + U_2 \ .
\end{equation}
As in Sect.~\ref{sec:eee}, we first expanded $U_k$ in elliptical elements. 
To the first order in the mass ratios, zeroth order in the eccentricities, and second order in the inclinations, we have
\begin{equation}
    \begin{split}
    U_k =
     - & k_{2} \frac{\mathcal{G}  m_k^2 \Ru^5 }{4 a_k^3 a_k'^3}\Bigg[1+3\cos (2 \theta-2 \theta'-2 \lambda_k+2 \lambda_k') \\
    &-\frac{3}{2} \inc_k^2 \, \bigg(1-\cos (2 \lambda_k-2 \Omega_k)
    +\cos (2 \theta-2 \theta'-2 \lambda_k+2 \lambda_k') \\& 
    \quad \quad -\cos (2 \theta-2 \theta'+2 \lambda_k'-2 \Omega_k) \bigg) \\
    &-\frac{3}{2} \inc_k'^2 \, \bigg(1-\cos (2 \lambda_k'-2 \Omega_k')
    +\cos (2 \theta-2 \theta'-2 \lambda_k+2 \lambda_k') \\&
    \quad \quad -\cos (2 \theta-2 \theta'-2 \lambda_k+2 \Omega_k')\bigg)\\
    &+3 \inc_k \inc_k' \, \bigg( \cos (\theta-\theta'-2 \lambda_k+2 \lambda_k'+\Omega_k-\Omega_k')\\
        &\quad \quad +\cos (\theta-\theta'-\Omega_k+\Omega_k') \phantom{\bigg)} \\
    &\quad \quad - \cos (\theta-\theta'-2 \lambda_k+\Omega_k+\Omega_k') \phantom{\bigg)} \\
    &\quad \quad- \cos (\theta-\theta'+2 \lambda_k'-\Omega_k-\Omega_k')
    \bigg)\Bigg] \ .
    \end{split} 
\end{equation}
We first performed the canonical change of variables that uses the resonant angles (Eq.\,(\ref{canonic_var})), and then changed to the complex Cartesian coordinates (Eq.\,(\ref{eq:complex_cartesian_coordinates})), to get
\begin{equation}
    \begin{split}
        U_k= - & k_{2} \frac{\mathcal{G} m_k^2 \Ru^5  \beta_k^{12} \mu_k^6}{4 \Gamma_k^7 \Gamma_k'^7} \, \Bigg[\Gamma_k \Gamma_k'-3\Gamma_k' \, \ybk \yk-3\Gamma_k \, \ybk' \yk'\\
        &- 6 p_k \, \bigg(\Gamma_k' \, \yi \ybi+\Gamma_k \, \yi' \ybi'+\Gamma_k' \, \yj \ybj+\Gamma_k \, \yj' \ybj'\bigg)\\
        &+3 \, \bigg( \Gamma_k \Gamma_k'-\Gamma_k' \, \ybk \yk-\Gamma_k \, \ybk' \yk'\\
        &\quad-6 p_k (\Gamma_k' \, \yi \ybi+\Gamma_k \, \yi' \ybi'+\Gamma_k' \, \yj \ybj+\Gamma_k \, \yj' \ybj') \bigg) \\
        &\quad \quad \times \cos (2 (\vartheta-\vartheta'-q_k (\gamma -\gamma' )))\\
        &+\frac{3}{2} \Gamma_k \, \bigg(\ybk'^2+\yk'^2\bigg) \cos (2 q_k \gamma')\\
        &+\frac{3}{2}\Gamma_k' \, \bigg(\ybk^2+\yk^2\bigg) \cos (2 q_k \gamma)\\
        &+3 \sqrt{\Gamma_k \Gamma_k'} \, \bigg(\ybk \yk'+\ybk' \yk\bigg)\cos (\vartheta-\vartheta')\\
        &-3 \sqrt{\Gamma_k \Gamma_k'} \, \bigg(\ybk \ybk'+\yk \yk'\bigg) \cos (\vartheta-\vartheta'+2 q_k \gamma')\\
        &-3 \sqrt{\Gamma_k \Gamma_k'} \, \bigg(\ybk \ybk'+\yk \yk'\bigg) \cos (\vartheta-\vartheta'-2 q_k \gamma)\\
        &+3 \sqrt{\Gamma_k \Gamma_k'} \, \bigg(\ybk \yk'+\ybk' \yk\bigg) \cos (\vartheta-\vartheta'-2 q_k (\gamma -\gamma' ))\\
        &+\frac{3}{2}\Gamma_k' \, \bigg(\ybk^2+\yk^2\bigg) \cos (2 (\vartheta-\vartheta'+q_k \gamma'))\\
        &+\frac{3}{2}\Gamma_k \, \bigg(\ybk'^2+\yk'^2\bigg) \cos (2 (\vartheta-\vartheta'-q_k \gamma))\\
        &-\frac{3}{2} \ii \, \Gamma_k \, \bigg(\ybk'^2-\yk'^2\bigg) \sin (2 q_k \gamma')\\
        &-\frac{3}{2} \ii \, \Gamma_k' \, \bigg(\ybk^2-\yk^2\bigg) \sin (2 q_k \gamma)\\
        &+3 \ii \sqrt{\Gamma_k \Gamma_k'} \, \bigg(\ybk' \yk-\ybk \yk' \bigg)\sin (\vartheta-\vartheta') \\
        &+3 \ii \sqrt{\Gamma_k \Gamma_k'} \, \bigg(\ybk \ybk'-\yk \yk' \bigg) \sin (\vartheta-\vartheta'+2 q_k \gamma')\\
        &-3 \ii \sqrt{\Gamma_k \Gamma_k'} \, \bigg(\ybk \ybk'-\yk \yk' \bigg) \sin (\vartheta-\vartheta'-2 q_k \gamma)\\
        &-3 \ii \sqrt{\Gamma_k \Gamma_k'} \, \bigg(\ybk' \yk-\ybk \yk' \bigg) \sin (\vartheta-\vartheta'-2 q_k (\gamma -\gamma' ))\\
        &-\frac{3}{2} \ii \, \Gamma_k' \, \bigg(\ybk^2-\yk^2 \bigg) \sin (2 (\vartheta-\vartheta'+q_k \gamma'))\\
        &+\frac{3}{2} \ii \, \Gamma_k \, \bigg(\ybk'^2-\yk'^2 \bigg) \sin (2 (\vartheta-\vartheta'-q_k \gamma))\Bigg] \ ,
    \end{split}
\end{equation}
with $p_1=-p/2$, $p_2=1+p/2$, $q_1=1+2/p$, and $q_2=1$.

We note that $\sigma$ does not appear in the expression of $U_k$.
Therefore, in the presence of tides the parameter $\Sigma$ (Eq.\,(\ref{SigmaTOT})) remains conserved.
The fast angle $\gamma$ is still present in the expression of $U_k$, but at this stage we cannot perform an average as in Sect.~\ref{cqaa} because $\gamma'$ is considered as a parameter that can later cancel with $\gamma$ (see Eq.\,(\ref{eq:gamma'_expansion})).

\subsection{Secular equations of motion}
\label{seomt}

The equations of motion are obtained from Eq.\,(\ref{tidalH}) using the Hamilton equations.
The additional contributions from tides only derive from the tidal potential energy $U_k$, and are given by
\begin{equation}
\label{tidal:eom}
\dot y_k = \ii \sum_{j=1}^2 \frac{\partial U_j}{\partial \ybk} \ , \quad
\dot \Gamma = - \sum_{j=1}^2 \frac{\partial U_j}{\partial\gamma} \ , \quad
\dot \Theta =-  \sum_{j=1}^2 \frac{\partial U_j}{\partial\vartheta} \ .
\end{equation}
In principle, we should also write the equations for $\dot \gamma$ and $\dot \vartheta$, but these angles disappear from the equations of motion with some of the following simplifications, and so we do not need them to get a closed set for the secular evolution of the system.

To handle the expression of the primed quantities, we need to use a tidal model.
For simplicity, we adopt here the weak friction model \citep[eg.][]{Singer_1968, Alexander_1973}, which assumes a constant and small time delay, $\dtU$.
This model is widely used and provides very simple expressions for the tidal interactions, because it can be made linear \citep[eg.][]{Mignard_1979}:
\begin{equation}\label{linear_model}
\lambda_k' \approx \lambda_k - n_k \dtU \quad \mathrm{and} \quad
\theta' \approx \theta - \angrotU \dtU \ .
\end{equation}
It follows for the remaining primed quantities that
\begin{equation}\label{eq:y_k'_expansion}
    \yk' \approx \yk - \ii \yk \left( p_2 \mmj + p_1 \mmi\right)\dtU \ ,
\end{equation}
\begin{equation}\label{eq:gamma'_expansion}
    \gamma' \approx \gamma - p_1 \left(\mmj-\mmi\right)\dtU \ , 
\end{equation} 
\begin{equation}\label{eq:vartheta'_expansion}
    \vartheta' \approx \vartheta + \left( p_2 \mmj + p_1 \mmi - \angrotU \right)\dtU \ ,
\end{equation}
with
\begin{equation}
\overline{n}_k = \beta_k^3 \mu_k^2 / \Gamma_k^{3} \ .
\end{equation}
We then substituted expressions (\ref{eq:y_k'_expansion}) to (\ref{eq:vartheta'_expansion}) into the equations of motion (\ref{tidal:eom}) and averaged over the fast angle $\gamma$ (as in Eq.\,(\ref{averHam})) to finally get the secular equations for the tidal evolution
\begin{equation}\label{eq:tidal_y1_equation_of_motion}
    \dot{y}_{1}=
    -\frac{3}{2}\frac{\mathcal{D}_{1}}{\Gamma_1^{13}}\left(2\ii p+\mmi \dtU\right) \yi
    +3\ii (p+2)\frac{\mathcal{D}_{2}}{\Gamma_2^{13}}\,\yi \ ,
\end{equation}
\begin{equation}\label{eq:tidal_y2_equation_of_motion}
    \dot{y}_{2}=
    -\frac{3}{2}\frac{\mathcal{D}_{2}}{\Gamma_2^{13}}\left(-2\ii(p+2)+\mmj\dtU\right)\yj
    -3\ii p\frac{\mathcal{D}_{1}}{\Gamma_1^{13}}\,\yj \ ,
\end{equation}
\begin{equation}\label{eq:tidal_Gamma_equation_of_motion}
\begin{split}
     \dot{\Gamma}=3\frac{\mathcal{D}_{1}}{\Gamma_1^{13}}&\left(1+2/p\right)\bigg[\Big(\Gamma_1 + 6p(\yi\ybi+\yj\ybj) -\yi\ybi \Big)\,\angrotU\\
     &-\Big(\Gamma_1+6p(\yi\ybi+\yj\ybj)\Big)\,\mmi\bigg] \, \dtU \\
     +3\frac{\mathcal{D}_{2}}{\Gamma_2^{13}}&\bigg[\Big(\Gamma_2-6(p+2)(\yi\ybi+\yj\ybj)-\yj\ybj\Big)\,\angrotU\\
     &-\Big(\Gamma_2-6(p+2)(\yi\ybi+\yj\ybj)\Big)\,\mmj\bigg]\,\dtU \ ,
     \end{split}
\end{equation}

\begin{equation}\label{eq:tidal_Theta0_equation_of_motion}
\begin{split}
     \dot{\Theta}=-3\frac{\mathcal{D}_{1}}{\Gamma_1^{13}}&\bigg[\Big(\Gamma_1+6p(\yi\ybi+\yj\ybj)-\yi\ybi\Big)\,\angrotU\\
     &-\Big(\Gamma_1+6p(\yi\ybi+\yj\ybj)-\yi\ybi\Big)\,n_1\bigg]\, \dtU \\
     -3\frac{\mathcal{D}_{2}}{\Gamma_2^{13}}&\bigg[\Big(\Gamma_2-6(p+2)(\yi\ybi+\yj\ybj)-\yj\ybj\Big)\,\angrotU\\
     &-\Big(\Gamma_2-6(p+2)(\yi\ybi+\yj\ybj)-\yj\ybj\Big)\,n_2\bigg] \, \dtU \ ,
     \end{split}
\end{equation}
where
\begin{equation}
    \mathcal{D}_{k}=k_2\mathcal{G}m_k^2\beta_k^{12}\mu_k^6\Ru^5 \ .
\end{equation}
We note that in the expressions of $\yi$ (Eq.\,(\ref{eq:tidal_y1_equation_of_motion})) and $\yj$ (Eq.\,{\ref{eq:tidal_y2_equation_of_motion})), we have a conservative contribution (imaginary terms) and a dissipative contribution (real terms in $\dtU$).
The conservative contributions result from a permanent tidal deformation and only slightly modify the fundamental frequencies of the system, while the dissipative contributions modify the secular evolution.

\subsection{Tidal constraints}\label{subsec:tidal_equations}

Tidal dissipation induces variations in the parameters $\Gamma$ (Eq.\,(\ref{eq:tidal_Gamma_equation_of_motion})) and $\Theta$ (Eq.\,(\ref{eq:tidal_Theta0_equation_of_motion})).
Then, the coefficients $\mathcal{K}$, $\mathcal{O}$, $\mathcal{S}$, and $\mathcal{R}$ appearing in the Hamiltonian (\ref{cartHam}) slowly change in time (Eqs.\,(\ref{Gb1})$-$(\ref{DeltaRef})), modifying the phase space (Fig.~\ref{fig:level_curves}), which translates into a secular evolution of the system.
We note that for the oblateness coefficients (Eqs.\,(\ref{Octerm}) and (\ref{Odterm})), changes are observed not only due to $\Gamma$, but also in $J_2$, according to \citep[eg.][]{Correia_Rodriguez_2013}
\begin{equation}
 J_2 = k_{\rm f} \frac{\angrotU^2\Ru^3}{3\mathcal{G}m_0} \ , 
\end{equation}
where $\angrotU = \Theta / C $ (Eq.\,(\ref{Lrot})) and $k_{\rm f}$ is the fluid second Love number for potential. 
Using the present rotation rate and $J_2$ of Uranus (Table \ref{table:physical_orbital_parameters}), we obtained $k_{\rm f}= 0.356 $.

\subsubsection{Orbital evolution}

For a better understanding of the tidal evolution of the satellites' orbits, we can compute the secular evolution of the semi-major axes and inclinations from Eqs.~(\ref{smak}) and (\ref{inckL}) as
\begin{equation}\label{eq:semi_major_axix_tidal_evolution}
   \dot{a}_k \approx 2 \At_k \left(\frac{\angrotU}{n_k} (1-\tfrac{1}{2} \inc_k^2 ) -1 \right) \, a_k \ ,
\end{equation}
\begin{equation}\label{eq:inclination_tidal_evolution}
 \dot \inc_k \approx -\frac{\At_k}{2} \frac{\angrotU}{n_k} \, \inc_k \ ,
\end{equation}
with
\begin{equation}\label{eq:Ktide}
\At_k = \frac{3 \mathcal{G}  m_k^2 \Ru^5}{\beta_k a_k^8} \, k_2 \dtU  \ .
\end{equation}
Both Ariel and Umbriel have very low inclinations (Table \ref{table:physical_orbital_parameters}). 
Combined with the fast rotation rate of Uranus ($\angrotU/n_k \gg 1$), we conclude that tides induce an outward migration of the satellites (Eq.\,(\ref{eq:semi_major_axix_tidal_evolution})), and they damp their inclinations (Eq.\,(\ref{eq:inclination_tidal_evolution})). 

\subsubsection{Resonance encounter}

\label{subsec:resonratios}

We can extrapolate the past evolution of the semi-major axes and determine when the 5/3~MMR encounter may have occurred (Eq.\,(\ref{eq:semi_major_axix_tidal_evolution})).
Neglecting the effect of the inclination and assuming a constant rotation rate for Uranus ($\angrotU/n_k \approx cte$), we get
\begin{equation}\label{int:sma_back}
   \frac{a_2^7 d a_2}{m_2\left(\angrotU / n_1 -1 \right)}  \approx
   \frac{a_1^7 d a_1}{m_1\left(\angrotU / n_2 -1 \right)}  \ ,
\end{equation}
which can be integrated to obtain $a_2$ as a function of $a_1$.
Starting with the present system (Table \ref{table:physical_orbital_parameters}) and assuming that the satellites were not temporarily captured into resonance, we can move backwards until the ratio of the nominal resonance is achieved (Eq.\,(\ref{keplerreson}))
\begin{equation}\label{nominal:sma_back}
 a_2 / a_1 \approx \left(5/3\right)^{2/3}  \approx 1.4057 \ .
\end{equation}
This allowed us to get the best estimate for the semi-major axes of Ariel and Umbriel at the exact 5/3~MMR (Fig.~\ref{fig:MMR_crossing}),
\begin{equation}\label{nominal:sma}
 a_1/\Ru = 7.3906 \ , \quad a_2/\Ru = 10.3891 \ ,
\end{equation}
and to compute other related parameters, such as $\Gamma$ (Eq.\,(\ref{Gammafix})).

\begin{figure}
    \centering
    \includegraphics[width=\linewidth]{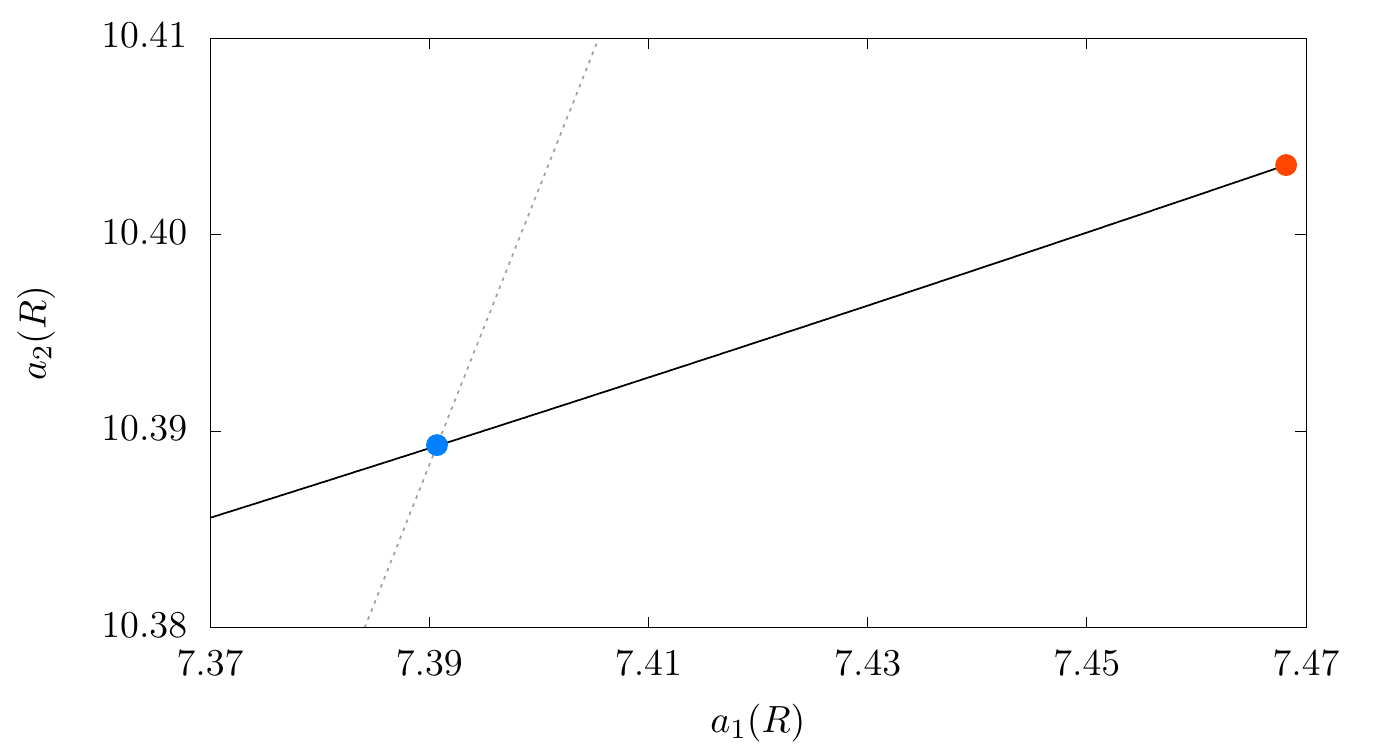}
    \caption{Tidal evolution of the semi-major axes of Ariel and Umbriel (Eq.\,(\ref{int:sma_back})). The orange point marks the current observed values (\mbox{Table \ref{table:physical_orbital_parameters}}). The dashed line gives the position of the nominal 5/3~MMR relation between $a_2$ and $a_1$ (Eq.(\,\ref{nominal:sma_back})). The blue point gives the best estimation for the semi-major axes at the nominal resonance (Eq.\,(\ref{nominal:sma})).} 
   \label{fig:MMR_crossing}
\end{figure}

\subsubsection{Evolution timescale}

The orbital evolution timescale depends on $\dtU$, which is related to the tidal dissipation (Eq.\,(\ref{linear_model})).
A more commonly used dimensionless quantity to measure the tidal dissipation is given by the quality factor \citep[eg.][]{Correia_Valente_2022}, 
\begin{equation}\label{eqQfactor}
    Q = 1 / (2\angrotU\dtU) \ .
\end{equation}

As a result of studying the likelihood of resonance crossing in the Uranian system, \citet{Tittemore_Wisdom_1990} stated that the 2/1~MMR Ariel-Umbriel cannot be crossed, while the 3/1~MMR Miranda-Umbriel can be, and thus they constrained the interval of $Q$ to be between $11\,000$ and $39\,000$. 
Following the same approach, but using more recent measures for the satellites' masses \citep{Jacobson_2014}, we find that $Q=8\,000$ is a more suitable value.
From Eq.\,(\ref{eqQfactor}), we then computed $\dtU \approx 0.62$~s.
In fact, the exact dissipation rate depends on the product $k_2 \dtU$ (Eq.\,(\ref{eq:Ktide})).
As in previous studies, in the present work we adopted $k_2 = 0.104$ \citep{Gavrilov_Zharkov_1977}, which translates into
\begin{equation}\label{k2sQ}
    \frac{k_2}{Q} = \num{1.3e-5} \quad \Leftrightarrow \quad k_2 \dtU = 0.064~\mathrm{s} \ .
\end{equation}

Using Eq.\,(\ref{eq:semi_major_axix_tidal_evolution}), we estimate that the 5/3~MMR was crossed about 640~Myr ago.
We can also estimate the damping timescale for the inclinations of both satellites (Eq.\,(\ref{eq:inclination_tidal_evolution})).
We get
\begin{equation}
\label{damp:inc}
    \tau_{\rm inc} \approx \frac{4 \beta_k a_k^8 n_k Q}{3 \mathcal{G}  m_k^2 \Ru^5 k_2} \ ,
\end{equation}
which yields about 180~Gyr and 1500~Gyr for Ariel and Umbriel, respectively.
We hence conclude that the presently observed inclination values are likely unchanged since the system crossed the 5/3~MMR.

\subsection{Capture probability}
\label{theoretical:capprob}

The behaviour of the system when it crosses the 5/3~MMR is not straightforward, since the problem has two degrees of freedom.
As explained in Sect.~\ref{sec:poincare}, the different behaviours while crossing the resonance depend on the energy of the system, which corresponds to different inclinations.
In order to get an idea of the critical inclinations that tend to trap the system in resonance or skip it, we can build a simplified one degree-of-freedom model.

The system can either be in resonance with the angle $\psii$ or $\psij$. 
Following \citet{Tittemore_Wisdom_1988}, we retained only the terms associated with each angle and obtained their associated one degree-of-freedom simplified Hamiltonian (Eq.\,(\ref{cartHam}))
\begin{equation}\label{simpleHam}
    \mathcal{H}_k=\Ak \, \yk\ybk+ \Bk \, (\yk\ybk)^2+\frac{\Ck}{2}  \, (\yk^2+\ybk^2) \ ,
\end{equation}
with
\begin{equation}
    \mathcal{A}_1 = \Ka + \Oc + \Sa + \Se \ , \quad \mathcal{B}_1 = \Kb \ , \quad \mathcal{C}_1 = \Rd \ ,
\end{equation}
\begin{equation}
    \mathcal{A}_2 = \Ka + \Od + \Sa + \Sf \ , \quad \mathcal{B}_2 = \Kb \ , \quad \mathcal{C}_2 = \Re \ .
\end{equation}

In this simplified case, we could then directly apply the theory of the adiabatic invariant for resonance capture \citep{Henrard_1982, Henrard_Lemaitre_1983}. 
In general, the phase space of a given resonance presents three different regions delimited by the separatrix (see for instance Fig.~\ref{fig:level_curves}c): a libration region, with area $\Jk^R$, and two circulation regions, one outside the libration region and another encircled by the separatrix, with an area $\Jk^C$.
Consider an initial trajectory with some energy $\Ham_k < \Ham_0$. 
As the system evolves due to tides, the phase space and the areas of each region also change.
When the trajectory encounters the resonance, depending on the exact place where the separatrix is crossed, the system can either evolve into the libration region (resonance) or into the inner circulation region.
We can compute the capture probability analytically, provided that the evolution is adiabatic, that is, the tidal induced variations are much slower than the conservative inclination variations.
The capture probability can then be obtained by the modification of the phase space with time, that is, by the change in the areas encircled by the separatrix \citep[eg.][]{Yoder_1979b, Henrard_1982}:
\begin{equation}\label{eq:p_C}
P_\mathrm{cap} = \frac{\dot {\Jk^R} }{ \dot {\Jk^R} + \dot {\Jk^C} } \ .
\end{equation}

The area of the resonance region is obtained by integrating over the separatrix,
\begin{equation}
    \Jk^R = \ii \oint \yk \, d\ybk \ .
\end{equation}
The energy of the separatrix ($\HamS$) can be found by finding the hyperbolic points (see Sect.~\ref{subsec:equilibrium_points}) and computing the Hamiltonian (\ref{simpleHam}) at these points,
\begin{equation}\label{eq:separatrix_energy}
   \HamS =-\frac{(\Ak+\Ck)^2}{4 \Bk} \ .
\end{equation}
Replacing $\HamS$ into expression (\ref{simpleHam}), we find, for the separatrix points, that
\begin{equation}\label{eq:y_k_bar_y_k}
    \yk =-\frac{2 \Ak \ybk  \pm \sqrt{-2 \Ck \left(\Ak+2 \ybk^2 \Bk +\Ck\right)^2 / \Bk}}{4 \ybk^2 \Bk +2 \Ck} \ .
\end{equation}
There are two solutions, the $+$ corresponding to the branch between the libration and the outer circulation region ($\Jk^+$), and the $-$ corresponding to the branch between the libration and the inner circulation region ($\Jk^-$).
Therefore, we have $\Jk^R = \Jk^+ + \Jk^-$ and $\Jk^C = - \Jk^- $, which gives (Eq.\,(\ref{eq:p_C}))

\begin{equation}\label{eq:capture_probility_equation}
    P_\mathrm{cap} = \frac{\dot {\Jk^+} + \dot {\Jk^-} }{ \dot {\Jk^+} } = 1 + \frac{\partial \Jk^-}{\partial\Gamma} \bigg/ \, \frac{\partial \Jk^+}{\partial\Gamma} \ ,
\end{equation}
where,
\begin{equation}
    \begin{split}
        \frac{\partial \Jk^-}{\partial\Gamma}&=\frac{1}{\Bk^2}\left(\arcsin{\left(\sqrt{-\frac{\Ck}{\Ak}}\right)}+\frac{\pi }{2}\right)(\Bk \Ak'-\Ak \Bk')\\
        &+\frac{1}{\Bk^2}\sqrt{-\frac{\Ck+\Ak}{\Ck}} (\Bk \Ck'-\Ck \Bk')   \ ,
    \end{split}
\end{equation}

\begin{equation}
    \begin{split}
        \frac{\partial \Jk^+}{\partial\Gamma}&=\frac{1}{\Bk^2}\left(\arcsin{\left(\sqrt{-\frac{\Ck}{\Ak}}\right)}-\frac{\pi }{2}\right)(\Bk \Ak' - \Ak \Bk')\\
        &+\frac{1}{\Bk^2}\sqrt{-\frac{\Ck+\Ak}{\Ck}} (\Bk \Ck'-\Ck \Bk') \ , 
    \end{split}
\end{equation}
and $\Ak' = {\partial \Ak}/{\partial\Gamma}$, $\Bk' = {\partial \Bk}/{\partial\Gamma}$, and $\Ck' = {\partial \Ck}/{\partial\Gamma}$. We note that 
since $\Gamma$ is the only time-dependent quantity appearing in $\Ak$, $\Bk$, and $\Ck$ (appendix~\ref{sec:Conservative_Hamiltonian_terms}), $\dot{\Jk}= \dot{\Gamma} \, {\partial\Jk} / {\partial\Gamma}$ (we neglected the small changes in $\Theta$ from the oblateness coefficients, ${\cal O}_k$).

\begin{figure}
    \centering
    \includegraphics[width=\linewidth]{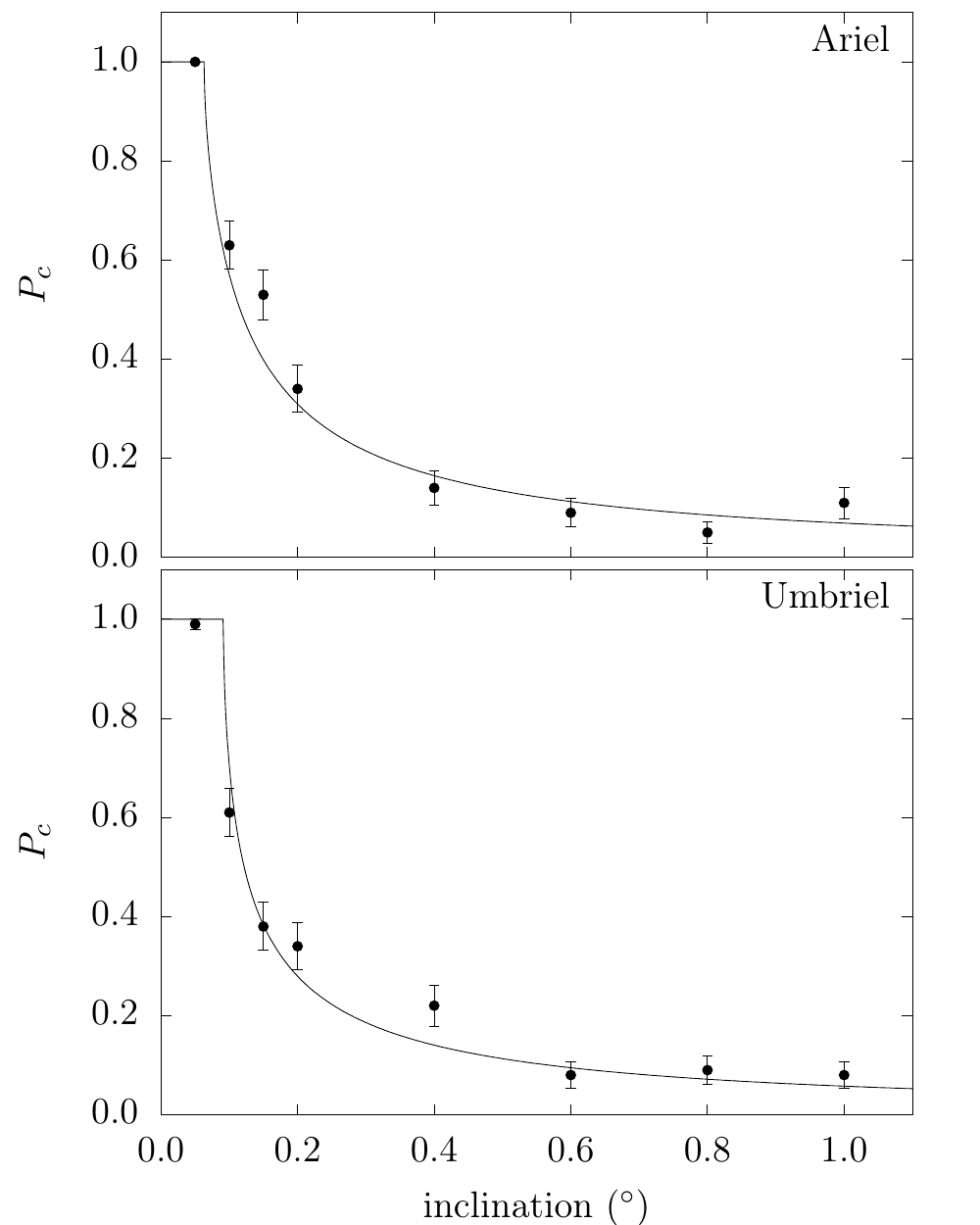}
    \caption{Capture probabilities in the $\psii$ (top) and $\psij$ (bottom) resonances. The solid line gives the theoretical approximation given by expression (\ref{eq:capture_probility_equation}), while the dots give the results of numerical simulations. We ran 100 initial conditions with $\psia_k$ differing by $1.8\degree$.}
    \label{fig:capture_probability_iAU}
\end{figure}

In Fig.~\ref{fig:capture_probability_iAU}, we show the probability of capture in the $\psii$ and $\psij$ resonances obtained with expression (\ref{eq:capture_probility_equation}). 
For some inclination values, we also show the results obtained with numerical integrations of the equations of motion derived from the simplified Hamiltonian (Eq.\,(\ref{simpleHam})) together with the secular tidal equations (Eqs.\,(\ref{eq:tidal_y1_equation_of_motion})$-$(\ref{eq:tidal_Theta0_equation_of_motion})).
For each initial inclination, we ran 100 simulations where the initial angle $\psia_k$ was uniformly sampled.
The amount of simulations captured in resonance at the end of the simulation are marked with a dot.
The statistical fluctuation, represented as error bars,  were estimated using binomial statistics, with the expression
\begin{equation}
    \Delta P = \sqrt{\frac{P_\mathrm{cap} }{N}\left(1-P_\mathrm{cap} \right)} \,,
\end{equation}
where $N=100$ is the number of simulations \cite[e.g.][]{Tittemore_Wisdom_1988}.
We observe that there is a good agreement between the theoretical curve (Eq.\,(\ref{eq:capture_probility_equation})) and the output of the numerical simulations, that is, the adiabatic approximation holds.

In Fig.~\ref{fig:capture_probability_iAU}, we observe that for initial inclinations lower than $~0.05\degree$, the system is always captured in resonance.
However, as we increase the initial inclination, the capture probability quickly decreases, it becomes $\sim 50\%$ for $\inc_k=0.1\degree$, and drops to $\sim 30\%$ for $\inc_k=0.2\degree$.
These results suggest that a system with nearly coplanar orbits cannot escape the 5/3~MMR, but for inclinations higher than about $0.1\degree$, it may be able to evade it.

We cannot completely rely on the conclusions obtained with the simplified Hamiltonian, mainly for two reasons.
One is because the complete Hamiltonian (Eq.\,(\ref{cartHam})) depends on the inclination of the other body.
When we simplified the Hamiltonian (Eq.\,(\ref{simpleHam})) for $\yi$, we dropped all terms in $\yj$ (and vice versa), which is equivalent to setting $\yj=0$.
However, if we set $\yj \ne 0$, more terms appear in the Hamiltonian, leading to a different distribution in the capture probabilities.
The other reason is that the complete Hamiltonian has two degrees of freedom, and so for some combinations of the inclination values, the system can be chaotic (see Sect.~\ref{sec:poincare}).
For the trajectories crossing the chaotic regions, the final outcome is unpredictable.

\section{Numerical simulations}\label{sec:Numerical_integration}

We have seen that Ariel and Umbriel almost certainly encountered the 5/3~MMR at some time in the past.
Depending on the inclination values of these satellites, the system may experience rather different behaviours.
Near coplanar orbits are expected to become trapped in resonance, while some inclination can lead to alternative scenarios.
In order to get a global view of the resonant dynamics with tides, in this section we integrate the complete set of differential equations (\ref{eq:conservative_motion_equations1}), (\ref{eq:conservative_motion_equations2}), and (\ref{eq:tidal_y1_equation_of_motion}) to (\ref{eq:tidal_Theta0_equation_of_motion}).  

\subsection{Setup}

When a resonance is crossed, we cannot perform a backward integration because we have a stochastic evolution.
Therefore, we need to place the system slightly before the resonance encounter and then integrate it forwards.
It is not possible to determine the exact semi-major axes prior to resonance crossing, but if the system does not spend much time in resonance, the semi-major axes should not differ much from the estimation given in Sect.~\ref{subsec:resonratios}.
We still need to slightly decrease $a_1$ (or increase $a_2$) to move the system out of the nominal resonance (Eq.\,(\ref{nominal:sma})).
Since tides are stronger in Ariel, we opted to shift $a_1$ and kept $a_2$ constant:
\begin{equation}\label{preres:sma}
 a_1/\Ru = 7.3892 \ , \quad a_2/\Ru = 10.3891 \ .
\end{equation}
These values of the semi-major axes allowed us to compute the initial $\Gamma$ parameter (Eq.\,(\ref{canonic_var})). 
For some given initial inclination values, we could also compute the initial value of $\Delta$ (Eq.\,(\ref{DeltaRef})), which translates into an initial $\delta > 0$ (Eq.\,(\ref{delta:equation})).

The physical properties of Uranus and its satellites can be found in Table \ref{table:physical_orbital_parameters}. 
The total angular momentum is conserved and can also be obtained from the present system,
\begin{equation}\label{Sigma:present}
\Sigma = \num{9.367247e-10} \ \mathrm{M_\odot \, au^2 \, yr^{-1}} \ .
\end{equation}
The initial rotational rate, $\angrotU$, was obtained from $\Sigma$ using Eq.\,(\ref{SigmaTOT}) with the pre-resonance semi-major axes (Eq.\,(\ref{preres:sma})).
Finally, for the tidal dissipation, we adopted $k_2 \dtU = 0.064~\mathrm{s}$ (Eq.\,(\ref{k2sQ})).

\subsection{Comparison with analytical estimations}

In general, tidal effects are weak and only correspond to small perturbations of the conservative dynamics (adiabatic approximation).
To verify that the numerical integrations do follow the theoretical predictions from Sects.~\ref{sec:Resonance_dynamics} and \ref{sec:Tidal_evolution}, we initially ran a few simulations with the full equations of motion and then superimposed the output in the equilibria map as a function of $\delta$ (Fig.~\ref{Fig:equilibrium_points}).
Since $\angrotU > n_k$, tidal effects are expected to increase the value of $\Gamma$ (Eq.\,(\ref{eq:tidal_Gamma_equation_of_motion})) and thus decrease the value of $\delta$ (Eq.\,(\ref{delta:equation})).
Therefore, the results of the simulations as a function of time must be read from the right to the left in the figure.

In Fig.~\ref{fig:overlap_low_inclination}, we show the results of a first experiment with very small initial inclinations for both satellites, $\inc_1 = \inc_2 = 0.01\degree$.
Initially, when $\delta > 0$, the system is in circulation with a small amplitude around the equilibrium point at zero $(\yi=0, \yj=0)$.
The system encounters the resonance when $\delta \approx 0$.
However, just before $\delta=0$, two stable equilibrium points emerge (corresponding to the 5/3~MMR), and the equilibrium point at zero becomes unstable.
The system is thus forced to follow one of the two resonance branches. 
As the system evolves and $\delta < 0$, the equilibrium point at zero becomes stable again.
However, because the amplitude of oscillation is small, the system closely follows the resonant branch, which corresponds to an increase in the inclinations.
We thus confirm that for initially near coplanar orbits, the system cannot avoid capture in the 5/3~MMR (Sect.~\ref{theoretical:capprob}).

\begin{figure}
    \centering
    \includegraphics[width=\linewidth]{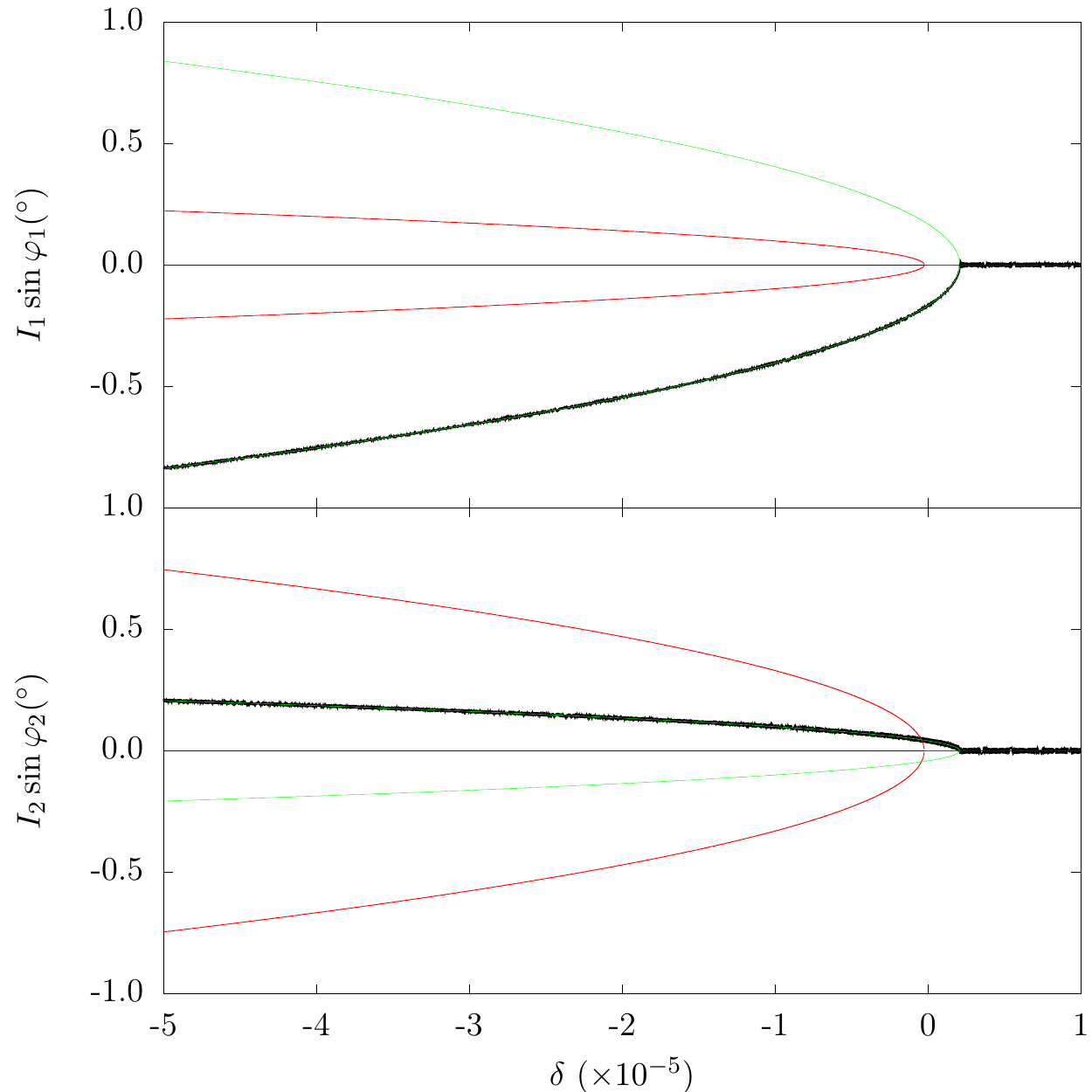}
    \caption{Tidal evolution of the system as function of $\delta$ for initial inclinations $\inc_1=\inc_2=0.01\degree$. The results of the numerical simulation (in black) are superimposed in the equilibria map (Fig.~\ref{Fig:equilibrium_points}). We show the evolution for the resonant angle $\psii$ (top) and $\psij$ (bottom).}
    \label{fig:overlap_low_inclination}
\end{figure}

\begin{figure*}
    \centering
    \includegraphics[width=\textwidth]{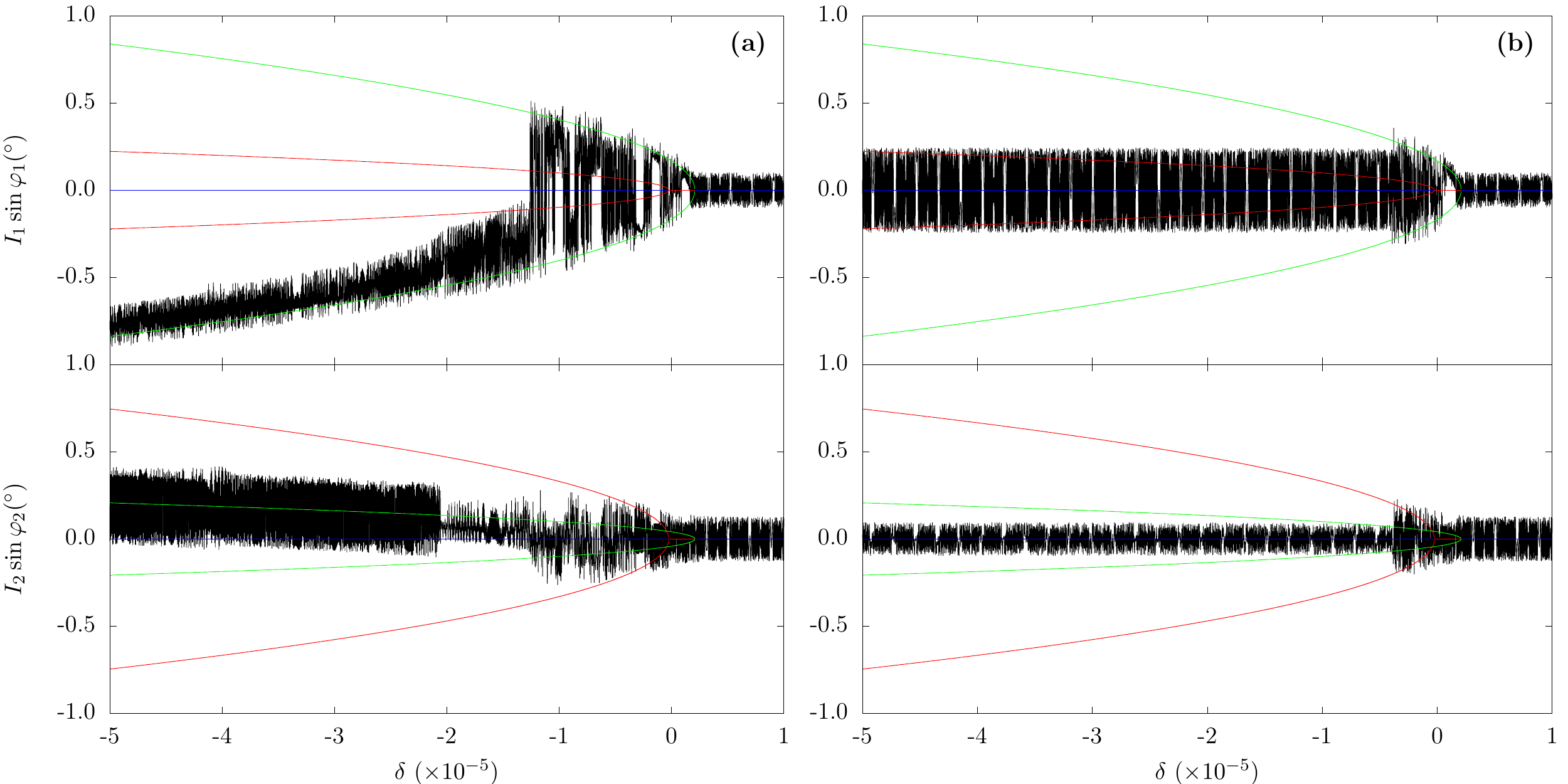}     
    \caption{Two examples of tidal evolution of the system as a function of $\delta$ for initial inclinations $\inc_1=\inc_2=0.1\degree$. The results of the numerical simulation (in black) are superimposed in the equilibria map (Fig.~\ref{Fig:equilibrium_points}). We show the evolution for the resonant angle $\psii$ (top) and $\psij$ (bottom).}
    \label{fig:overlap_both}
\end{figure*}
In Fig.~\ref{fig:overlap_both}, we show the results of a second experiment with higher initial inclinations for both satellites, $\inc_1 = \inc_2 = 0.1\degree$.
The initial evolution for $\delta > 0$ is similar to the case with lower initial inclinations (Fig.~\ref{fig:overlap_low_inclination}), except that the amplitude of oscillation is ten times larger in this case (corresponding to a system with a higher energy).
As the system encounters the resonance at $\delta \approx 0$, it is not able to follow one of the resonant equilibria and it remains in a chaotic region around the separatrix (Fig.~\ref{fig:Poincare_surface_ariel}f).
In the example on the left (Fig.~\ref{fig:overlap_both}a), after some time in the chaotic region with $\delta < 0$, the system finds a way into the libration region and follows one of the resonant branches in a quasi-periodic orbit (Fig.~\ref{fig:Poincare_surface_ariel}i).
On the other hand, in the example on the right (Fig.~\ref{fig:overlap_both}b), the system finds an alternative path back into the circulation region around the equilibrium point at zero in a quasi-periodic orbit (Fig.~\ref{fig:Poincare_surface_ariel}k).
We thus confirm that for initial orbits with some inclination, the system experiences a chaotic regime for some time (Sect.~\ref{sec:poincare}), after which it can either be captured in the 5/3~MMR, or escape it with some probability (Sect.~\ref{theoretical:capprob}).

\subsection{Impact of the initial inclination}

The present mean inclinations of Ariel and Umbriel (\mbox{Table \ref{table:physical_orbital_parameters}}) are likely unchanged after the system quits the 5/3~MMR (Eq.\,(\ref{damp:inc})). 
However, they may have been rather different before this encounter, as the inclinations undergo some excitations while crossing the resonance (Sect.~\ref{sec:Resonance_dynamics}).
The fact that the system is not trapped in resonance at present, implies that the separatrix was crossed at some point and the inclinations had to experience some chaotic oscillations (Fig.~\ref{fig:overlap_both}b).
Therefore, it is impossible to simply integrate backwards and determine the exact inclination values prior to the resonance encounter.
To have a more clear idea of what the system may have been, we need to perform a statistical study of its past evolution, starting with arbitrary initial inclinations for both Ariel and Umbriel and then reject those that are not coherent with the present observations.

For that purpose, we explored a mesh of initial inclinations ranging between $0.001\degree$ and $0.2\degree$ with a stepsize of $0.05\degree$.
For each pair ($\inc_1,\inc_2$), we ran 1abe000 simulations evenly sampled over the angle $\sigma$ (Eq.\,(\ref{eq:resonace_argument})) for 100~Myr, in a total of 25\,000 experiments.
In Table \ref{tab:simulations_set}, we list the complete set of initial conditions together with a summary of the outcome of the resonance crossing.
The results of the simulations are presented in percentage because they can be seen as a statistical distribution of a given final evolution possibility.
The total number of events of each kind can be simply obtained multiplying the percent by ten.

\begin{table*}[]
\caption{Initial conditions and summary of the numerical simulations's results of the 5/3~MMR crossing. }\label{tab:simulations_set}
\begin{center}                        
    \begin{tabular}{c|c|c|c|c|c|c c|c c}
       \#  & $\inc_1 \, (\degree)$ & $\inc_2 \, (\degree)$ & $P_c (\%)$ & $P_e (\%)$ & $P_s (\%)$ & $\Rmeani \, (\degree)$ & $\Rsigma_1 \, (\degree)$&  $\Rmeanj \, (\degree)$  & $\Rsigma_2 \, (\degree)$\\
       \hline\hline
       1 & $0.001$ & $0.001$ & $100.0$ & $-$ & $-$ &  & & &\\
       2 & $0.05$ & $0.001$ & $100.0$ & $-$ & $-$ & & & & \\
       3 & $0.10$ & $0.001$ & $99.2$ & $0.8 $ & $-$ & & & & \\
       4 & $0.15$ & $0.001$ & $100.0$ & $-$ & $-$ & & & & \\
       5 & $0.20$ & $0.001$ & $100.0$ & $-$ & $-$ & & & & \\
       \hline
       6 & $0.001$ & $0.05$ & $100.0$ & $-$ & $-$ & & & & \\
       7 & $0.05$ & $0.05$ & $100.0$ & $-$ & $-$ & & & & \\
       8 & $0.10$ & $0.05$ & $97.3$ & $2.5$ & $0.2$ & & & & \\
       9 & $0.15$ & $0.05$ & $82.4$ & $10.5$ & $7.1$ & $0.149$ & $0.062$ & $0.056$ & $0.023$ \\
       10 & $0.20$ & $0.05$ & $80.9$ & $9.3$ & $9.8$ & $0.187$ & $0.054$ & $0.047$ & $0.019$ \\
       \hline
       11 & $0.001$ & $0.10$ & $90.1$ & $9.2$ & $0.7$ & $0.170$ & $0.041$ & $0.047$ & $0.016$ \\
       12 & $0.05$ & $0.10$ & $83.1$ & $13.6$ & $3.3$ & $0.167$ & $0.050$ & $0.041$ & $0.022$ \\
       13 & $0.10$ & $0.10$ & $50.6$ & $28.5$ & $20.9$ & $0.145$ & $0.059$ & $0.049$ & $0.020$ \\
       14 & $0.15$ & $0.10$ & $38.8$ & $19.2$ & $42.0$ & $0.119$ & $0.068$ & $0.068$ & $0.020$ \\
       15 & $0.20$ & $0.10$ & $36.2$ & $15.8$ & $48.0$ & $0.141$ & $0.066$ & $0.081$ & $0.023$ \\
       \hline
       16 & $0.001$ & $0.15$ & $39.6$ & $26.5$ & $33.9$ & $0.112$ & $0.068$ & $0.083$ & $0.019$ \\
       17 & $ 0.05$ & $0.15$ & $38.6$ & $25.8$ & $35.6$ & $0.114$ & $0.069$ & $0.095$ & $0.024$ \\
       18 & $0.10$ & $0.15$ & $35.1$ & $25.1$ & $39.8$ & $0.088$ & $0.081$ & $0.109$ & $0.017$ \\
       19 & $0.15$ & $0.15$ & $31.0$ & $23.2$ & $45.8$ & $0.099$ & $0.082$ & $0.127$ & $0.023$ \\
       20 & $0.20$ & $0.15$ & $30.0$ & $15.9$ & $54.1$ & $0.104$ & $0.085$ & $0.136$ & $0.021$ \\
       \hline
       21 & $0.001$ & $0.20$ & $32.7$ & $26.2$ & $41.1$ & $0.107$ & $0.073$ & $0.144$ & $0.018$ \\
       22 & $0.05$ & $0.20$ & $31.9$ & $28.6$ & $39.5$ & $0.116$ & $0.070$ & $0.152$ & $0.018$ \\
       23 & $0.10$ & $0.20$ & $30.1$ & $23.9$ & $46.0$ & $0.109$ & $0.072$ & $0.162$ & $0.015$ \\
       24 & $0.15$ & $0.20$ & $27.1$ & $22.1$ & $50.8$ & $0.117$ & $0.075$ & $0.173$ & $0.013$ \\
       25 & $0.20$ & $0.20$ & $27.0$ & $15.2$ & $57.8$ & $0.119$ & $0.076$ & $0.182$ & $0.014$ \\ \hline
    \end{tabular}
    \end{center}
    $\inc_k$ is the initial inclination of each satellite; $P_c$ is the number of simulations trapped in resonance more than 10 Myr; $P_e$ is the number of simulations that were captured in resonance, but escaped in less than 10 Myr; $P_s$ is the number of simulations that skipped the resonance; and $\langle\inc_k\rangle$ and $\Rsigma_k$ are the mean and the standard deviation, respectively, of a Rice distribution adjusted to the final results.
\end{table*}

\subsubsection{Capture probability}

For each run, we first evaluated whether capture in resonance occurred or not.
Capture takes place whenever at least one resonant angle, $\psia_k$ (Eq.\,(\ref{psiiang})), switches from circulation to libration, and the mean motion ratio becomes approximately constant ($n_1/n_2 \approx 5/3$).
The mean motion ratio criterion is usually more useful to automatically detect captures  because inside chaotic regions the behaviour of the resonant angles can be quite erratic.
Conversely, capture does not occur when all resonant angles remain in circulation, which introduces only a small jump in the mean motion ratios.
For this last case, we consider that the resonance is skipped.
In Table \ref{tab:simulations_set}, the relative number of these events is dubbed as $P_s$ (skip probability).

The capture probability can be simply evaluated as $1-P_s$.
However, when a capture occurs, we still have to distinguish between trajectories that remain captured for long periods of time from those that are able to escape shortly after.
Indeed, in many resonant lockings, the inclinations remain chaotic and the system is able to evade the resonance after some time.
In theory, all trajectories evading the resonance could lead to the present system. 
In practice, that is not possible, because the longer the system stays in resonance, the higher the final inclinations become (Fig.~\ref{fig:overlap_low_inclination}), and they cannot be conciliated with the presently observed values (Table \ref{table:physical_orbital_parameters}).
Moreover, \citet{Cuk_etal_2020} have shown that as long as Ariel and Umbriel stay in resonance, the inclinations of the three other large satellites of Uranus also grow, in particular that of Miranda.
After only 10~Myr in resonance, is it impossible to explain the observed values since tides are unable to efficiently damp the inclinations (Eq.\,(\ref{damp:inc})).
Therefore, in our analysis, we split the capture events into those that stay in resonance for more than 10~Myr and those that are able to evade it before that time.
In Table \ref{tab:simulations_set}, the relative number of escapes is dubbed as $P_e$ (escape probability) and the relative number of long-term entrapments is referred to as $P_c$ (capture for more than 10~Myr probability).

In most sets of the simulations shown in Table \ref{tab:simulations_set}, all three scenarios described above (capture, escape, and skip) are simultaneously present.
To better illustrate the different possibilities, in Fig.~\ref{fig:simulation_examples}, we show one example of each case, corresponding to set \#17 (with an  initial $\inc_1 = 0.05\degree$ and $\inc_2 = 0.15\degree$).

\begin{figure*}
    \centering
    \includegraphics[width=\textwidth]{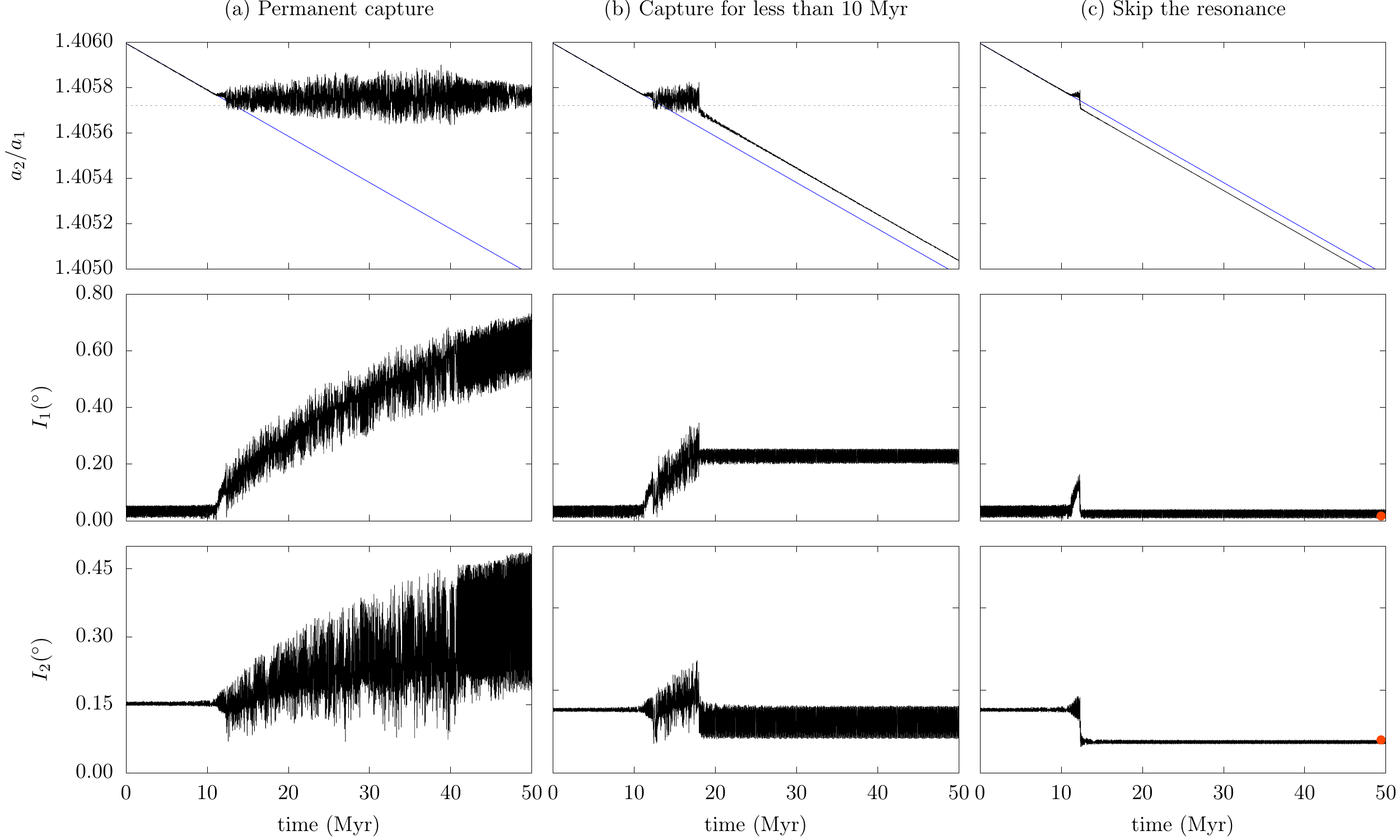}
    \caption{Three examples of a system crossing the 5/3~MMR with initial $\inc_1 = 0.05\degree$ and $\inc_2 = 0.15\degree$ (set \#17 in Table \ref{tab:simulations_set}). We show the ratio between the semi-major axes (top), the inclination of Ariel (middle), and the inclination of Umbriel (bottom) as a function of time. Each column corresponds to a different simulation. We show an example of a system that is permanently caught in resonance (a), one that is captured but evades the resonance in less than 10~Myr (b), and another that skips the resonance without capture (c). The blue line gives the asymptotic evolution predicted by Eq.\,(\ref{eq:semi_major_axix_tidal_evolution}), while the dashed line gives the position of the nominal resonance Eq.\,(\ref{nominal:sma_back})). The orange dots mark the present mean inclinations (Table \ref{table:physical_orbital_parameters}).}
    \label{fig:simulation_examples}
\end{figure*}

In Fig.~\ref{fig:simulation_examples}a, the system is permanently captured in resonance.
Prior to the resonance encounter, the semi-major axes ratio follows the asymptotic evolution predicted by Eq.\,(\ref{eq:semi_major_axix_tidal_evolution}).
When the system comes across the resonance (at $t\approx 11$~Myr), the semi-major axes ratio becomes constant (since $n_1/n_2 \approx 5/3$), deviating considerably from the asymptotic line.
Indeed, looking at Fig.~\ref{fig:MMR_crossing}, when capture occurs, instead of following the black curve, the system follows the dashed line, corresponding to the nominal resonance (Eq.\,(\ref{nominal:sma_back})).
Shortly after being captured, the system enters into the chaotic region.
The inclinations of both satellites are excited and grow steadily on average, in particular that of Ariel, which oscillates between the two resonant branches (green lines in Fig.\,\ref{Fig:equilibrium_points}).
For a time $t\approx 40$~Myr, the system finds a way into the libration region and becomes quasi-periodic, as in the example shown in Fig.~\ref{fig:overlap_both}a.

In Fig.~\ref{fig:simulation_examples}b, the system is temporarily captured in resonance for less than 10~Myr.
As in the previous example, the semi-major axes ratio initially follows the asymptotic evolution predicted by Eq.\,(\ref{eq:semi_major_axix_tidal_evolution}), but it switches to a constant ratio as soon as capture in resonance occurs (at $t\approx 11$~Myr).
As before, during the resonant entrapment, the system enters into the chaotic region and the inclinations of both satellites start to grow.
However, for a time around $t \approx 18$~Myr (that is, just 7~Myr after being captured), the system finds a way outside the chaotic region that breaks the resonant locking and returns into the circulation region around the equilibrium point at zero (blue line in Fig.~\ref{Fig:equilibrium_points}).
From that point on, the semi-major axes ratio again follows the asymptotic evolution predicted by Eq.\,(\ref{eq:semi_major_axix_tidal_evolution}), though restarting with a slightly higher ratio, and the inclinations of both satellites stabilise around a given constant mean value.
The final inclination of Ariel is always higher than its initial value because of the forcing during the resonant phase. 
The final inclination of Umbriel is less impacted by this mechanism and it does not have a systematic trend because it oscillates around the initial value with a large amplitude during the resonance crossing.

Finally, in Fig.~\ref{fig:simulation_examples}c, the system shortly skips the resonance.
As usual, the semi-major axes ratio initially follows the asymptotic evolution predicted by Eq.\,(\ref{eq:semi_major_axix_tidal_evolution}), and it undergoes some perturbations during the resonance crossing. However, it quickly returns to the asymptotic evolution, though restarting with a slightly lower ratio.
In this case, there is almost no chaotic evolution for the inclinations because, just after crossing the separatrix, the system directly goes into the circulation region around the equilibrium point at zero (blue line in Fig.~\ref{Fig:equilibrium_points}).
The inclination of Ariel initially briefly grows since, owing to the topology of the 5/3~MMR (Fig.~\ref{fig:level_curves}), it is not possible to reach the inner circulation region without moving around the resonant equilibrium for a short time (less than 1~Myr).
Since in the example \#17 the initial inclination of Ariel is indeed relatively small, for a short moment the inclination of Ariel follows the resonant branch and has to increase (green line in Fig.\,\ref{Fig:equilibrium_points}).
Nevertheless, as soon as the equilibrium point at zero becomes stable, the system moves back into the circulation region and the inclination of Ariel drops.
In the example shown in Fig.~\ref{fig:simulation_examples}c, the inclination of Ariel decreases to a mean value smaller than the initial one, but actually in other examples it can be anything between zero and the maximum previously attained.
The brief resonant excitation only involves the angle $\psii$, and so it does not impact the inclination of Umbriel much.
Its amplitude grows due to the mutual interactions with Ariel, but the mean value remains constant.
However, as the resonance is skipped, the mean value of Umbriel inclination suddenly drops to a lower level.
This reduction is always observed because the inner circulation region is confined within a region of low inclination for Umbriel, in particular for $\delta$ values very close to zero (Fig.~\ref{fig:level_curves}c).

Fig.~\ref{fig:simulation_examples}c also provides an example of a simulation where the system crosses the 5/3~MMR and subsequently evolves into the presently observed configuration (orange dots, taken from \mbox{Table \ref{table:physical_orbital_parameters}}).
Not all simulations \#17 that skip the resonance end in the present system, though a significant number is consistent with it.
On the other hand, some simulations starting with different initial conditions (Table \ref{tab:simulations_set}) can also evolve into the present state.
Therefore, only a more refined analysis of the final distribution of the inclinations of both Ariel and Umbriel can provide more insight into the occurrence likelihood of the currently observed system.

\subsubsection{Statistics}

For the set of simulations $\#1-\#8$ (Table \ref{tab:simulations_set}), that is to say those with initial $\inc_2 \le 0.05\degree$, the system is captured in resonance nearly in $100\%$ of the cases.
For the initial inclination of Umbriel, these results are thus in agreement with the predictions of the simplified model presented in Sect.~\ref{theoretical:capprob}. 
Since capture in the 5/3~MMR is not observed today, we can immediately exclude this range of initial conditions.
We hence conclude that, regardless of the initial inclination of Ariel, the system requires some moderate initial inclination for Umbriel ($\inc_2 \gtrsim 0.1\degree$) to evolve into the present state.

\begin{figure*}
    \centering
    \includegraphics[width=\textwidth]{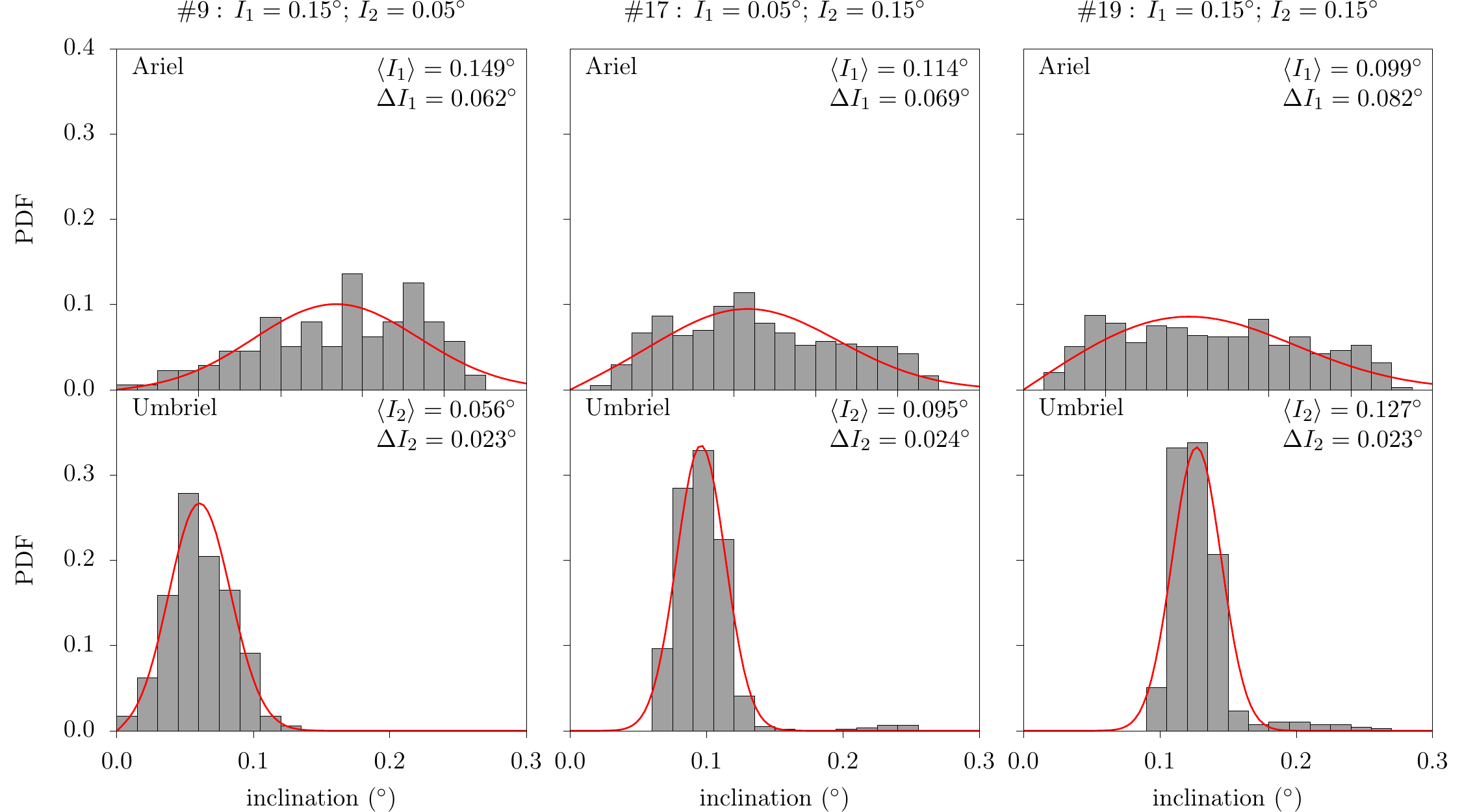} 
    \caption{Histograms for the final distribution of the inclinations of Ariel (top) and Umbriel (bottom) for different sets of initial conditions. We also plotted the best fit Rice distribution (Eq.\,(\ref{Rice:dist:f})) to each histogram (red curve) and the corresponding mean inclination, $\Rmean$, and standard deviation, $\Rsigma$.
    We used the sets \#9 (left), \#17 (centre), and \#19 (right) from Table \ref{tab:simulations_set}.}
    \label{fig:histograms_rice}
\end{figure*}

The resonance crossing is a stochastic process and therefore the same initial inclinations with a slightly different initial resonant angle may end up in a completely different final state.
In all of the other sets of simulations that we performed ($\#9-\#25$), the system can either remain captured for a long time, escape the resonance in less than 10~Myr, or simply skip it. 
In general, as we increase the initial inclinations of both satellites, the probability of capture in resonance decreases, in conformity with the analysis from Sect.~\ref{theoretical:capprob}.
Interestingly, the number of trajectories temporarily captured for less than 10~Myr does not change much with the initial conditions, they occur around $20\% - 25\%$ of the time.
This suggests that the chaotic diffusion, which characterises this transient regime, is not very sensitive to the initial inclinations.
Finally, as a result of the previous two outcomes, as we increase the initial inclinations of both satellites, the number of simulations that simply skip the resonance also increases.

For those systems that skip or escape the 5/3~MMR, one can ask if the final inclinations are in agreement with the present observations.
However, there is not an easy answer because of the chaotic diffusion.
Indeed, the final inclinations of Ariel and Umbriel never end exactly with the same values, but they rather follow some kind of statistical distribution.
Considering only the trajectories that quickly evade or skip the resonance ($P_e$ and $P_s$), we can build a histogram to better understand how they are distributed for each set of simulations $\#9-\#25$ (i.e. we only consider simulations with less than $95\%$ of capture probability).

In Fig.~\ref{fig:histograms_rice}, we show three examples of histograms for the sets $\#9$ ($\inc_1 = 0.15\degree, \inc_2 = 0.05\degree$), $\#17$ ($\inc_1 = 0.05\degree, \inc_2 = 0.15\degree$), and $\#19$ ($\inc_1 = 0.15\degree, \inc_2 = 0.15\degree$).
The final inclinations of Ariel and Umbriel are distributed in classes with a size of $0.015\degree$ and the number of events in each class is normalised by the total number of trajectories that quickly evaded or skipped the resonance.
We observe that for Ariel, the final inclinations are more or less evenly distributed between $0.01\degree$ and $0.25\degree$, while for Umbriel the final inclinations closely pile up around some mean value.

To better analyse the results in a systematic way, and since $\inc_k \propto | \yk | $ (Eq.\,(\ref{yinc})), we fitted a Rice distribution to each data set \citep{Rice_1945}, which describes the modulus of a random walk variable in two dimensions. 
This function has the form
\begin{equation}\label{Rice:dist:f}
    f (\inc)=\frac{\inc}{\Rsigma^2} B_0\left(\frac{\inc \, \Rmean}{\Rsigma^2}\right) \exp\left(-\frac{\inc^2+\Rmean^2}{2\Rsigma^2}\right) \ ,
\end{equation}
where $\Rmean$ is the mean inclination, $\Rsigma$ is the standard deviation, and $B_0(x)$ is the modified Bessel function, given by
\begin{equation}
    B_0\left(x\right)=\sum_{n=0}^\infty \frac{x^{2n}}{n!^2}  \ .
\end{equation}
In Fig.~\ref{fig:histograms_rice}, for each histogram we also show the curve of the Rice distribution corresponding to the best fit parameters $\Rmean$ and $\Rsigma$.
We verified there is a reasonably good agreement between the histogram and the derived distribution, in particular for the final inclination of Umbriel.
We hence adopted the best fit parameters obtained in this way to characterise each data set.

The statistical results obtained are listed in Table\ref{tab:simulations_set}.
From a detailed analysis, for the trajectories that are not trapped in resonance, we observe the following:
for Ariel, regardless of the its initial inclination value, the final inclinations are always more or less uniformly distributed\footnote{We note that by construction, the Rice distribution must be zero when the inclination is zero and it has to peak around the mean value (Eq.\,(\ref{Rice:dist:f})). As the final inclinations of Ariel are more or less uniformly distributed, a step function would provide a better adjustment. Nevertheless, we kept the Rice distribution for Ariel for simplicity since it is able to correctly capture the mean value and the dispersion around it.} in the interval $\inc_1 \in [0.01\degree,0.25\degree]$;
 and for Umbriel, the final inclinations are gathered around the mean value with a standard deviation $\Rsigma_2 \approx 0.02\degree$, and the mean value for the final inclination of Umbriel increases with the initial inclinations of both satellites.
A more subtle analysis additionally shows that the average final inclination of Umbriel is always lower than its initial value and it also depends on the initial inclination of Ariel. 
It is approximately given by (see Fig.~\ref{fig:adjust})
\begin{equation}\label{empiricdist}
   \Rmeanj - \inc_2 \approx   - 0.06\degree + 0.22 \, \inc_1 \ ,
\end{equation}
that is, to say the lower the initial inclination of Ariel is, the larger the decrease observed in the initial inclination of Umbriel.

\begin{figure}
    \centering
    \includegraphics[width=\columnwidth]{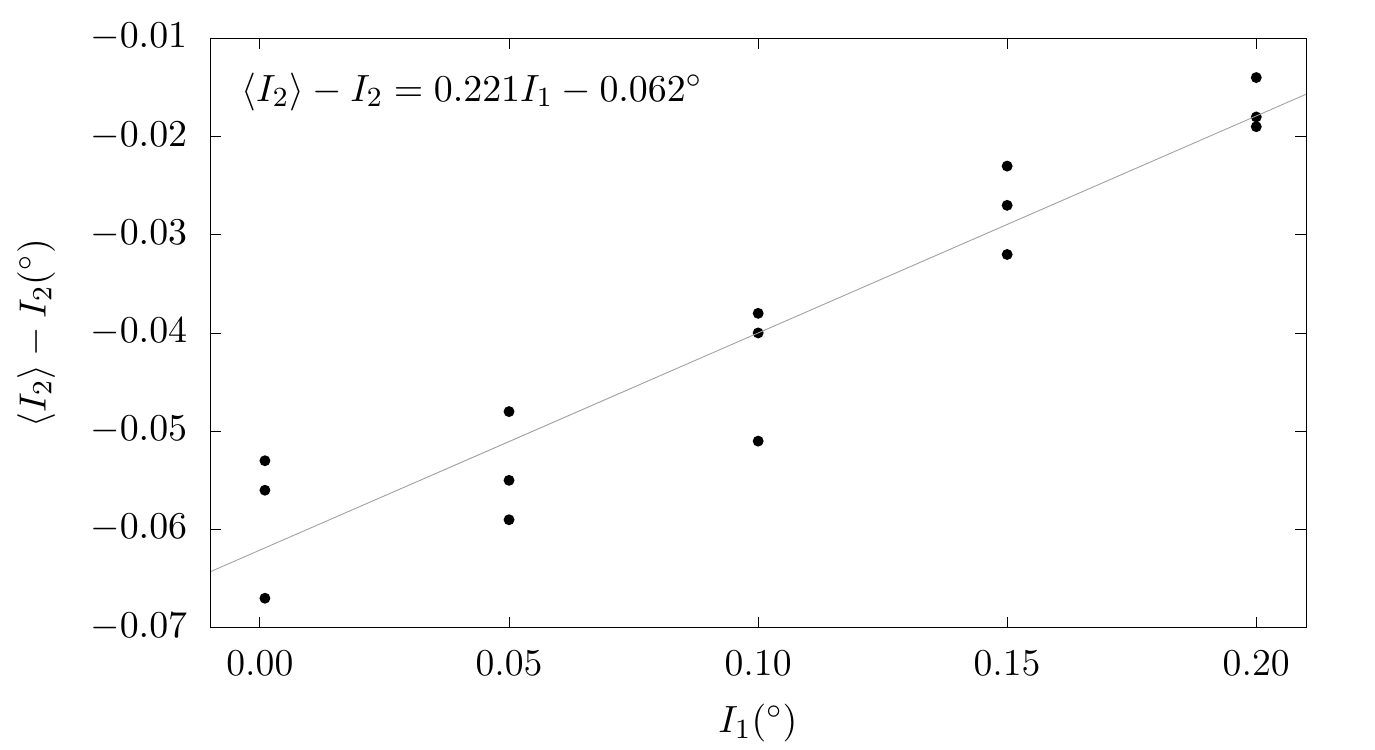} 
    \caption{Variation in the average inclination of Umbriel after crossing the 5/3~MMR as a function of the initial inclination of Ariel. The points are taken from the simulations \#11 to \#25 in Table \ref{tab:simulations_set}, and the line gives the adjustment of a linear regression.}
    \label{fig:adjust}
\end{figure}

The present average inclinations of Ariel and Umbriel are $\inc_1 \approx 0.02\degree$ and $\inc_2 \approx 0.08\degree$, respectively (Table \ref{table:physical_orbital_parameters}).
After crossing the 5/3~MMR, the inclination of Ariel is always approximately uniformly distributed and not very sensitive to the initial conditions.
The presently observed value is thus compatible with near zero or moderate initial inclinations of both satellites ($\inc_k < 0.2\degree$), but no additional constraints can be derived.
However, the final inclination of Umbriel strongly depends on the initial inclination of both satellites (Eq.\,(\ref{empiricdist})).
We hence conclude that the best configuration that reproduces the present data was obtained for initial $\inc_1 \approx 0.20\degree$ and $\inc_2 \approx 0.10\degree$ (\#15, Table \ref{tab:simulations_set}) or initial $\inc_1 \lesssim 0.05\degree$ and $\inc_2 \approx 0.15\degree$ (\#16 and \#17, Table \ref{tab:simulations_set}).

The satellites of Uranus were likely formed in a circumplanetary disk \citep[eg.][]{Pollack_etal_1991, Szulagyi_etal_2018, Ishizawa_etal_2019, Inderbitzi_etal_2020}, and so the initial inclinations should have been extremely small.
Therefore, initial inclinations of $0.15\degree$ or higher are very difficult to explain.
However, Umbriel was most likely previously involved in a 3/1~MMR with Miranda \citep{Tittemore_Wisdom_1989, Tittemore_Wisdom_1990}, which can excite the inclination of Umbriel to $0.15\degree$ prior to the encounter with the 5/3~MMR.
We hence conclude that the most likely scenario for the initial inclinations of Ariel and Umbriel is $\inc_1 \lesssim 0.05\degree$ and $\inc_2 \approx 0.15\degree$ (\#16 and \#17, Table \ref{tab:simulations_set}).

\section{Conclusion}\label{sec:Conclusion}

Ariel and Umbriel have almost certainly passed through the 5/3~MMR in the past owing to the tidal evolution of their orbits.
However, the exact mechanism that allows the system to evade capture in this resonance is a mystery.
For coplanar orbits (zero inclinations), the eccentricity of at least one satellite must be close to 0.01 at the time, which is unlikely because tides are expected to quickly damp the eccentricities to nearly zero \citep{Tittemore_Wisdom_1988}.
For non-coplanar orbits, the inclinations appear to grow to high values, which is unlikely because tides are very inefficient to damp the inclinations to the presently observed near zero values \citep{Cuk_etal_2020}.

To address this question, in this paper we revisited the 5/3~MMR crossing problem only focussing on the inclination.
We adopted a secular resonant two-satellite circular model with low inclinations, using a Hamiltonian approach similar to \citet{Tittemore_Wisdom_1989} for the study of the 3/1~resonance between Miranda and Umbriel.
However, in our model we included the spin evolution of Uranus, we used complex Cartesian coordinates, and we adopted the total angular momentum of the system as a canonical variable, which is conserved and naturally removes one degree of freedom from the problem.
We thus only needed to perform one average over a fast angle, instead of the widely used average over two fast angles \citep[eg.][]{Tittemore_Wisdom_1988, Tittemore_Wisdom_1989, Michtchenko_Ferraz-Mello_2001, Alves_etal_2016}.
We also implemented a Hamiltonian extension for tides based on the constant time-lag model, which provides the exact tidal evolution for all variables in the problem.

Our model is valid for any second order MMR, and thus not restricted to the 5/3~MMR between Ariel and Umbriel.
For instance, it can be directly applied to the passage through the 3/1~MMR between Miranda and Umbriel.
More generally, it can also be used to study a large number of planetary systems near second order resonances (in the circular approximation), such as the 3/1~MMR for HD\,60532 \citep{Laskar_Correia_2009}, the 5/3~MMR for HD\,33844 \citep{Wittenmyer_etal_2016}, the 7/5~MMR for HD\,41248 \citep{Jenkins_etal_2013}, or the 9/7~MMR for Kepler-29 \citep{Vissapragada_etal_2020}.

Applying our model to Ariel and Umbriel, 
we have shown that, prior to the 5/3~MMR encounter, the system is in circulation around an equilibrium point at the origin of the coordinates $(\yi=0, \yj=0)$.
As the system approaches the resonance, this equilibrium becomes unstable, while two other stable symmetrical equilibria appear, corresponding to libration in resonance.
Shortly after, the equilibrium at zero becomes stable again, corresponding to a new circulation region (see Fig.~\ref{fig:level_curves}}).
For initial very low inclination values (low energy), the system is thus forced to follow one of the resonant equilibria (capture).
However, for moderate inclinations (higher energy), the system may encounter the resonance when all the stable equilibrium possibilities are already available and directly go to the circulation region (skip).
Alternatively, the system enters into a chaotic regime and can subsequently evade into the circulation (escape) or the libration regions (capture).
The chaotic nature of the system as a function of the energy is clearly portrayed in a sequence of Poincar\'e surface sections obtained with the modified H\'enon method (Fig.~\ref{fig:Poincare_surface_ariel}).

The crossing of the 5/3~MMR is a stochastic process, and so we performed a large number of numerical simulations covering many different combinations for the initial inclinations of Ariel and Umbriel.
The results show that the initial inclination of Umbriel must have been higher than about $0.1\degree$ to avoid a permanent capture.
Moreover, in order to conciliate the output of the simulations with the presently observed system (Table \ref{table:physical_orbital_parameters}), we find that the optimal inclinations for the satellites prior to the resonance encounter are $\inc_1 \lesssim 0.05\degree$ for Ariel and $\inc_2 \approx 0.15\degree$ for Umbriel.
In this configuration, about $60\%$ of the simulations avoid capture in resonance and the final inclination distribution of Umbriel clusters around the present mean value of $0.08\degree$.
The final inclination distribution of Ariel spreads between $0.01\degree$ and $0.25\degree$ with a nearly equal probability, which includes the present mean value of $0.02\degree$.
The inclination damping timescales of $\sim 180$~Gyr for Ariel and $\sim 1500$~Gyr for Umbriel additionally suggest that the inclination values obtained just after crossing the resonance have nearly remained unchanged up to the present.

The satellites of Uranus were presumably formed in a circumplanetary disk, and so the primordial eccentricities and inclinations should have been extremely small.
Before the encounter with the 5/3~MMR, the Uranian satellites may have crossed other MMRs \citep[eg.][]{Peale_1988, Cuk_etal_2020}.
In general, these resonances excite the eccentricities and the inclinations of the bodies involved in the commensurability.
Tides are very efficient at damping the eccentricities, but not the inclinations.
Therefore, while the remnant eccentricities are quickly eroded, the inclinations are fossilised until the next resonant encounter.
For instance, the present high inclination of Miranda ($\sim 4.3\degree$) is usually explained after a passage through the 3/1~MMR between Miranda and Umbriel that occurred several billion years ago.
Interestingly, this resonance also involves Umbriel, whose inclination can therefore also be excited, though to a much lower value than that of Miranda \citep{Tittemore_Wisdom_1989, Tittemore_Wisdom_1990}.
As a consequence, prior to the encounter with the 5/3~MMR, it is reasonable to assume a near zero inclination for Ariel and an inclination around $0.15\degree$ for Umbriel.

The results obtained in this paper are very compelling, but they need to be taken with caution because our model is limited to circular orbits and two satellites. 
In a non-circular model, the eccentricity terms introduce other resonant angles that provide additional libration and chaotic regions to the problem \citep{Tittemore_Wisdom_1988}.
Indeed, for the 3/1~MMR between Miranda and Umbriel, it was shown that the coupling between the eccentricity and inclination resonances may cause significant variations in the eccentricity evolution of Miranda \citep{Tittemore_Wisdom_1990}.
Moreover, the presence of the remaining three large satellites or Uranus can also introduce three-body resonances that may further excite the eccentricities and the inclinations \citep{Cuk_etal_2020}.
Therefore, in order to fully understand how exactly the system evaded the 5/3~MMR between Ariel and Umbriel and subsequently settled into the present state, future work should take into account the effect of the eccentricities and from all five major satellites in the Uranian system.

\begin{acknowledgements}
This work was supported by grant
SFRH/BD/143371/2019,
and by projects
CFisUC (UIDB/04564/2020 and UIDP/04564/2020),
GRAVITY (PTDC/FIS-AST/7002/2020),
PHOBOS (POCI-01-0145-FEDER-029932), and
ENGAGE SKA (POCI-01-0145-FEDER-022217),
funded by COMPETE 2020 and FCT, Portugal.
We acknowledge the Laboratory for Advanced Computing at University of Coimbra (\href{https://www.uc.pt/lca}{https://www.uc.pt/lca}) for providing the resources to perform the numerical simulations.
\end{acknowledgements}

\bibliographystyle{aa}
\bibliography{bibliography.bib}

\begin{appendix}
\section{Conservative Hamiltonian coefficients}\label{sec:Conservative_Hamiltonian_terms}
We note that 

\begin{equation}
    \Ka= -\frac{p}{2}\frac{\mu_1^2\beta_1^3}{\Gamma_1^3}+\left(1+\frac{p}{2}\right)\frac{\mu_2^2\beta_2^3}{\Gamma_2^3}
\end{equation}

\begin{equation}
    \Kb= -\frac{3}{2}\left(\frac{p^2}{4}\frac{\mu_1^2\beta_1^3}{\Gamma_1^4}+\left(1+\frac{p}{2}\right)^2\frac{\mu_2^2\beta_2^3}{\Gamma_2^4}\right)
\end{equation}


\begin{equation}\label{Octerm}
    \Oc=\frac{3}{2}J_2R^2   \left(-(p-1)\frac{\mu_1^4\beta_1^7 }{\Gamma_1^7}+(p+2)\frac{\mu_2^4\beta_2^7}{\Gamma_2^7}\right)
\end{equation}
\begin{equation}\label{Odterm}
    \Od=\frac{3}{2}J_2R^2    \left(-p\frac{\mu_1^4\beta_1^7 }{\Gamma_1^7}+(p+3)\frac{\mu_2^4\beta_2^7 }{\Gamma_2^7}\right)
\end{equation}


\begin{equation}
    \begin{split}
    \Sa & = \frac{\mathcal{G}\mu_2m_1m_2\beta_2^2}{\Gamma_2^3} \Bigg( 1+\frac{p}{2} + \\ & \quad \quad \quad \frac{\mu_2\beta_2^2}{\mu_1\beta_1^2} \left[\left(1+\frac{p}{2}\right)\frac{\Gamma_1^2}{\Gamma_2^2}+\frac{p}{2}\frac{\Gamma_1}{\Gamma_2}\right] \frac{\partial}{\partial \alpha}
    \Bigg) \, b^{(0)}_{\frac{1}{2}}(\alpha_0)
    \end{split}
\end{equation}

\begin{equation}
    \Se=\frac{\mathcal{G}\mu_2m_1m_2\beta_2^2}{4\Gamma_1\Gamma_2^2}\alpha_0 b^{(1)}_{\frac{3}{2}}(\alpha_0)
\end{equation}
\begin{equation}
    \Sf=\frac{\mathcal{G}\mu_2m_1m_2\beta_2^2}{4\Gamma_2^3}\alpha_0 b^{(1)}_{\frac{3}{2}}(\alpha_0)
\end{equation}
\begin{equation}
    \Sg=-\frac{\mathcal{G}\mu_2m_1m_2\beta_2^2}{2\Gamma_2^2\sqrt{\Gamma_1\Gamma_2}}\alpha_0b^{(1)}_{\frac{3}{2}}(\alpha_0)
\end{equation}


\begin{equation}
    \Rd=-\frac{\mathcal{G}\mu_2m_1m_2\beta_2^2}{4 \Gamma_1 \Gamma_2^2} \alpha_0 b^{(p+1)}_{\frac{3}{2}}(\alpha_0)
\end{equation}
\begin{equation}
    \Re=-\frac{\mathcal{G}\mu_2m_1m_2\beta_2^2}{4 \Gamma_2^3} \alpha_0 b^{(p+1)}_{\frac{3}{2}}(\alpha_0)
\end{equation}
\begin{equation}
    \Rf=\frac{\mathcal{G}\mu_2m_1m_2\beta_2^2}{2\Gamma_2^2\sqrt{\Gamma_1\Gamma_2}}\alpha_0 b^{(p+1)}_{\frac{3}{2}}(\alpha_0)
\end{equation}
where $\alpha_0\approx0.7114$ corresponds to $\alpha =  a_1 / a_2 $ at the nominal resonance (Eq.\,(\ref{nominal:sma_back})).

\end{appendix}

\end{document}